\documentclass[twocol]{ametsocV6.1}

\usepackage{amsmath,amsfonts,amssymb,bm}
\usepackage{mathptmx}
\usepackage{newtxtext}
\usepackage{newtxmath}
\usepackage{comment}
\usepackage[normalem]{ulem}

\ifdraft\else
\PassOptionsToPackage{hyphens}{url}
\usepackage{hyperref}
\hypersetup{
	breaklinks,
	colorlinks=true,
	linkcolor=blue,
    urlcolor=Purple,
	citecolor=Purple,
 }
\fi

\newcommand{\sgn}{\mathrm{sgn}}
\newcommand{\degree}{^\circ}

\title{Intrinsically Episodic Antarctic Shelf Intrusions of Circumpolar Deep Water via Canyons}

\authors{Ellie Q.~Y.~Ong,\aff{a\correspondingauthor{Ellie~Q.~Y.~Ong, ellie.ong@unsw.edu.au}}
Edward~Doddridge,\aff{b}
Navid~C.~Constantinou,\aff{c} \ifdraft\else\\\fi
Andrew~McC.~Hogg,\aff{c} and
Matthew~H.~England\aff{d}}

\affiliation{
    \aff{a}{Climate Change Research Center, Australian Center for Excellence in Antarctic Science \& Australian Research Council Center of Excellence for Climate Extremes, University of New South Wales, Sydney, NSW, Australia}\\
    \aff{b}{Australian Antarctic Program Partnership, Institute for Marine and Antarctic Studies, University of Tasmania, nipaluna / Hobart, Tasmania, Australia}\\
    \aff{c}{Research School of Earth Sciences \& Australian Research Council Center of Excellence for Climate Extremes, Australian National University, Canberra, ACT, Australia}\\
    \aff{d}{Centre for Marine Science and Innovation (CMSI) \& ARC Australian Center for Excellence in Antarctic Science, University of New South Wales, Sydney, NSW, Australia}
}

\abstract{The structure of the Antarctic Slope Current at the continental shelf is crucial in governing the poleward transport of warm water. Canyons on the continental slope may provide a pathway for warm water to cross the slope current and intrude onto the continental shelf underneath ice shelves, which can increase rates of ice shelf melting, leading to reduced buttressing of ice shelves, accelerating glacial flow and hence increased sea level rise.  Observations and modelling studies of the Antarctic Slope Current and cross-shelf warm water intrusions are limited, particularly in the East Antarctica region. To explore this topic, an idealised configuration of the Antarctic Slope Current is developed, using an eddy-resolving isopycnal model that emulates the dynamics and topography of the East Antarctic sector. Warm water intrusions via canyons are found to occur in discrete episodes of large onshore flow induced by eddies, even in the absence of any temporal variability in external forcings, demonstrating the intrinsic nature of these intrusions to the slope current system. Canyon width is found to play a key role in modulating cross-shelf exchanges; warm water transport through narrower canyons is more irregular than transport through wider canyons. The intrinsically episodic cross-shelf transport is found to be driven by feedbacks between wind energy input and eddy generation in the Antarctic Slope Current. Improved understanding of the intrinsic variability of warm water intrusions can help guide future observational and modelling studies in the analysis of eddy impacts on Antarctic shelf circulation.}

\begin{document}

\maketitle

%
%
%

%








\section{Introduction}

The Antarctic Slope Current (ASC) is a westward flowing current around Antarctica, lying close to the coast over the continental slope \citep{Thompson2018TheClimate}. The ASC is closely coupled to the Antarctic Slope Front, which is characterized by steeply sloping isopycnals located over the Antarctic continental slope, spinning up a westward geostrophic current between the saline open ocean and the fresher Antarctic continental shelf. A key water mass of the saline open ocean is the Circumpolar Deep Water (CDW) \citep[e.g.][]{Heywood2014OceanSlope,Thompson2018TheClimate,Morrison2020WarmCanyons,Daae2020NecessarySea}. The poleward transport of this warm CDW is regulated by the slope current; depending on the local structure of the ASC, CDW can intrude on to the continental shelf, transporting heat poleward to the Antarctic shelf. This process can induce basal melt of ice shelves and increase glacial and ice sheet flow, resulting in global sea level rise \citep{Depoorter2013CalvingShelves, Hattermann2014Eddy-resolvingOcean,Herraiz-Borreguero2016BasalAntarctica,Rintoul2016OceanShelf, DeConto2016ContributionRise, Gudmundsson2019InstantaneousShelves}.

Given the significant impact Antarctic ice shelf melt can have globally, the extent to which CDW intrudes underneath the ice shelves of the East Antarctic Ice Sheet is a major concern. The East Antarctic Ice Sheet holds a total sea-level equivalent of $\sim50\,\mathrm{m}$, an order of magnitude larger than that of the West Antarctic Ice Sheet \citep{Stokes2022ResponseChange}. However, much less research has been conducted in the East Antarctic compared to the West Antarctic region \citep{Morlighem2020DeepSheet,Stokes2022ResponseChange}. The East Antarctic Ice Sheet has previously been thought to be stable, with landlocked sectors that are not directly impacted by CDW intrusions and past melt rates that have been lower than those in the West Antarctic \citep[e.g.][]{Paolo2015VolumeAccelerating, Stokes2022ResponseChange}. However, recent studies point towards a greater mass loss than previously predicted, especially in marine-based sectors of the East Antarctic Ice Sheet exposed to CDW intrusions onto the continental shelf (e.g. \citet{Rignot2019Four1979-2017,Stokes2022ResponseChange}), highlighting the vulnerability of the East Antarctic Ice Sheet to melting due to warm water intrusions.

The sloped isopycnals of the ASC in East Antarctica are generally thought to act as a barrier between CDW offshore and the continental shelf, however, the presence of canyons in the topography on the continental shelf may provide a pathway for CDW flow on to the shelf. Oceanic observations have shown the presence of warm CDW intrusions through canyons in East Antarctica  \citep{Rintoul2016OceanShelf,Nitsche2017BathymetricMargin,Silvano2018FresheningWater, Silvano2019SeasonalityGlacier,Hirano2020StrongAntarctica, Ribeiro2021WarmAntarctica, Herraiz-Borreguero2022PolewardSheet}. However, our physical understanding of these warm CDW intrusions via canyons is incomplete, as observations are scarce \citep[e.g.][]{Pena-Molino2016Direct113E, Herraiz-Borreguero2016BasalAntarctica} and modelling studies are also limited, as outlined below. Our understanding of canyons' effect on warm water intrusions is also limited by the limited bathymetric measurements around the East Antarctic Margin \citep{Nitsche2017BathymetricMargin, Silvano2019SeasonalityGlacier, McMahon2023SouthernShelf}, which show many discrepancies between observations and bathymetric products. There is, however, a better understanding of shelf intrusions via canyons outside of East Antarctica, where some previous work has investigated mechanisms inducing an onshore flow in the presence of a slope current. Possible drivers of onshore flow include steering of a shelf break jet following strong canyon topography \citep{Williams2001TheTopography}, mean flow-topographic interactions inducing a cyclonic circulation in the canyon \citep{St-Laurent2013OnShelves, Fennel1991ResponsesForcing} or wave-topographic interactions under a jet residing further offshore \citep{Zhang2011ShelfTopography,St-Laurent2013OnShelves}. Another study has also investigated the effect of a continental slope on slope current variability, but this did not include the effect of a canyon \citep{Stern2015InstabilityBreak}. Hence, limited understanding of the effect of canyons on cross-slope CDW transport in the East Antarctic continental margin is a key motivation behind this research.

Existing modelling studies do not generally take into account all the key factors that influence how CDW intrusions occur on the East Antarctic continental shelf. Namely, these studies do not generally resolve eddies, canyons and East Antarctica in the same model. Many regional modelling studies of the East Antarctic continental margin are not fully eddy-resolving \citep[e.g.][]{Gwyther2014SimulatedShelves, Gwyther2018IntrinsicShelf, Nakayama2021AntarcticGlacier}, even though eddies are crucial in governing the transport of CDW onto the continental shelf \citep{Nst2011EddySea, Thompson2014EddyCirculation, Stewart2015Eddy-mediatedBreak, BoeiraDias2023SensitivityResolution}. To resolve eddies at the Antarctic continental margin, an order $\sim500\textrm{m} - 1\textrm{km}$ resolution is required to resolve a continental shelf Rossby radius of deformation of $\sim 4\textrm{km}$ (which corresponds to a $1/48$th-degree resolution) \citep{St-Laurent2013OnShelves, Stewart2015Eddy-mediatedBreak}. Conversely, existing eddy-resolving studies do not generally investigate the East Antarctic region and the associated `fresh shelf regime', where strong easterly winds result in the poleward Ekman transport of cool, surface water, the incropping of density surfaces and a strong ASC (e.g. \citet{Thompson2018TheClimate}). Instead, eddy-resolving modelling studies, such as the ones by \citet{Daae2017OnSystem} and \cite{Liu2017ModelingAntarctica}, generally focus on `dense shelf regime' regions of the Antarctic margin, where dense shelf water formation leads to the formation of Antarctic Bottom Water (e.g. \citet{Williams2010Antarctic140149E,Ohshima2013AntarcticPolynya}). These dense shelf water formation regions have different circulation regimes and dynamics compared to most of the East Antarctic margin \citep{Darelius2014HydrographyAntarctica,Darelius2016ObservedWater, Daae2017OnSystem, Morrison2020WarmCanyons}; in particular, the dense shelf regime has an additional bottom-intensified slope current linked to dense shelf water export, which is independent of the surface intensified current of the fresh shelf regime \citep{Huneke2023DecouplingExport}. The only eddy-resolving modelling study of the fresh shelf regime is that by \citet{Liu2022Topography-MediatedAntarctica}, who investigate CDW intrusions via canyons focuses on the mechanisms of cross-shelf exchange in a single topographic configuration. \citet{Liu2022Topography-MediatedAntarctica} find that warm water intrusions are steered onshore by the bottom pressure torque when the CDW flow first interacts with the mouth of the trough, with topographic Rossby waves driving a high frequency variabiltiy of $\sim$ 1 month.  Currently, there are no eddy-resolving modelling studies investigating the effect of canyon width and geometry on the ASC and CDW intrusions in a fresh-shelf regime, hence our understanding of CDW intrusions under different canyon topographies remains poor. Addressing this question is therefore the focus of the present study.

Here, we explore warm CDW intrusions through canyons in an idealised channel configuration based on the East Antarctic continental margin, using a range of canyon configurations. We model the fresh shelf regime dominant around the East Antarctic margin using an idealised eddy-resolving primitive equation model with isopycnal coordinates. The isopycnal channel model is forced by wind stress at the surface, with restoring boundary conditions to the north. Using isopycnal coordinates allows us to model the stratification of the ASC with few vertical layers at a minimal computational cost, which makes it feasible to explore the parameter space of canyon geometries in eddy-resolving simulations. Further details about the isopycnal channel model setup are given in Section~\ref{sec:model}. In Section~\ref{sec:CDW_intrusions}, we investigate the intrinsically episodic intrusions of CDW onto the shelf in the idealised channel simulations, showing the effect of canyon geometry on the regularity of these intrusions, and that the episodic variability of intrusions is linked to an intrinsic temporal variability of the ASC itself. In Section~\ref{sec:ASC_variability}, we examine the emergent intrinsic temporal variability of the ASC and develop a simplified low-order model, comprising of two non-linear ordinary differential equations, that is able to reproduce episodic ASC variability based on energy exchanges between different reservoirs. In Section~\ref{sec:discussion}, we discuss the implications of our results and future work.

\begin{figure*}
    \centering
    \includegraphics[width = \textwidth]{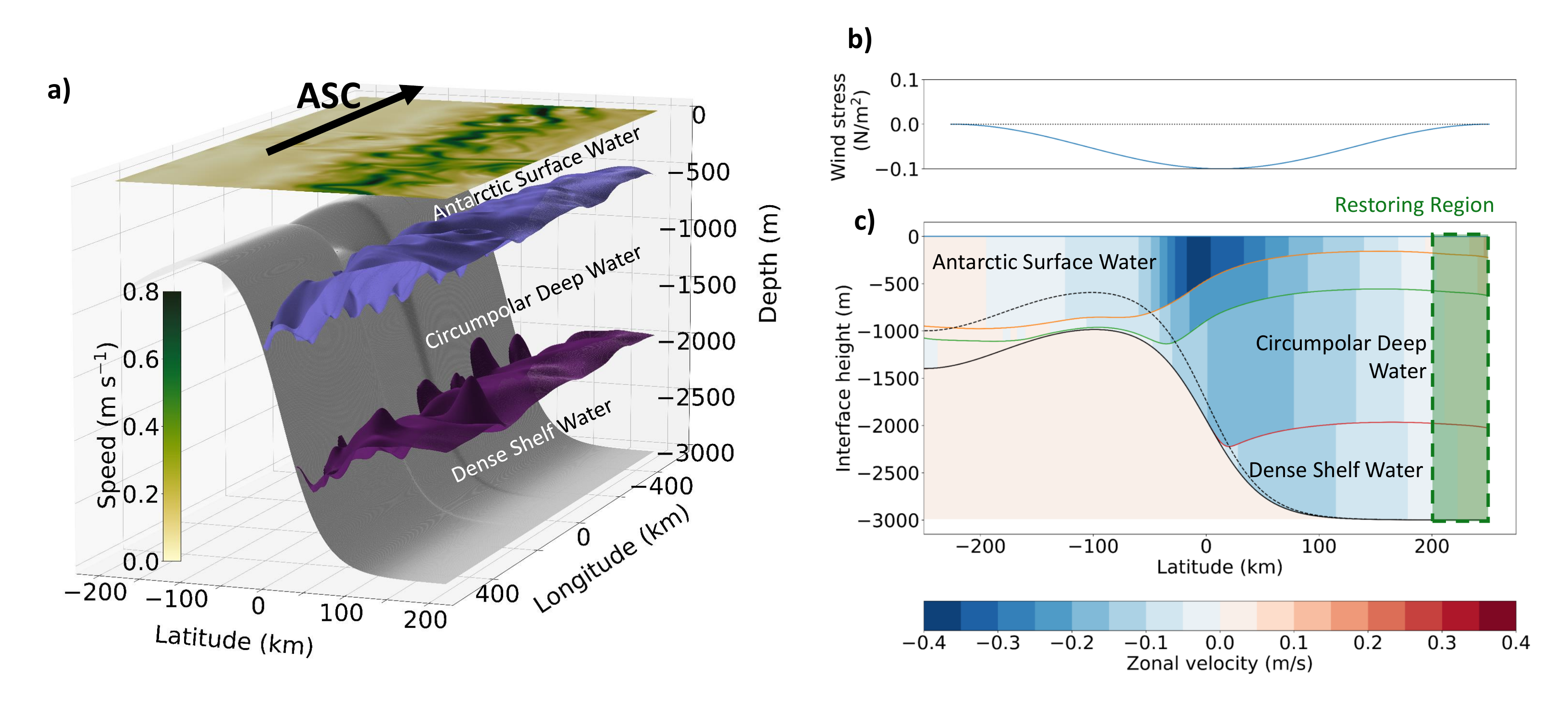}
    \caption{(a) Snapshot of isopycnal interfaces for the upper surface water, CDW and dense shelf water layers in the control simulation of the ASC, with the widest steep-sided canyon topography. Instantaneous surface speed is shown on the top isopycnal interface. (b) Profile of zonal wind stress forcing, applied at the surface. (c) Cross-sectional profile of isopycnal surfaces in the centre of the canyon between each of the four density layers: upper and lower surface water, CDW, and dense shelf water. Colours in each density layer represent the 10-year mean zonal velocity in each layer. Black dashed line shows the topography of the continental slope away from the canyon, while the black solid line shows the topography of the continental slope at the centre of the canyon. The restoring region where density interfaces are relaxed to a set height at the northern boundary, representing the open ocean, is shown in green. The densities of each isopycnal layer are, starting from the surface layer, $\rho_1 = 1027.8 \,\mathrm{kg\, m^{-3}}$, $\rho_2=1028.0 \,\mathrm{kg\, m^{-3}}$, $\rho_3=1028.1 \,\mathrm{kg\, m^{-3}}$, $\rho_4=1028.3 \,\mathrm{kg \, m^{-3}}$. }
    \label{fig:model_setup}
\end{figure*}

\section{Model setup}
\label{sec:model}

The primary experiments in this study utilize an isopycnal channel model of the ASC to investigate the effect of different canyon configurations on CDW intrusions. We also develop, and later introduce in Section \ref{sec:ASC_variability}, a low-order model of two ordinary differential equations that explains an emergent variability in ASC strength found in the isopycnal channel model. Section \ref{sec:model} will outline the setup of the primary isopycnal channel model, and the experiments that were run to understand CDW intrusions across different canyon configurations.  

The isopycnal channel model domain and forcings are designed to reproduce the fresh shelf regime \citep{Thompson2018TheClimate}, and are inspired by the configuration used by \citet{Constantinou2019EddyPerspective}. We use the Modular Ocean Model version~6 (MOM6) \citep{Adcroft2019TheFeatures} to solve the hydrostatic Boussinesq primitive equations in isopycnal coordinates. We have a zonally re-entrant channel on a beta-plane, with a zonal extent of $1000 \, \mathrm{km}$, meridional extent of $500 \, \mathrm{km}$, and maximum depth of $3\, \mathrm{km}$, that includes topography of a continental slope. The height of the continental slope, from the top of the sill on the edge of the continental shelf to the bottom, is $2.5\, \mathrm{km}$.  The Coriolis parameter is $f = f_0 + \beta y$, with $f_0 = -10^{-4} \, \mathrm{s}^{-1}$, $\beta = 1.5 \times 10^{-11} \, \mathrm{m}^{-1} \mathrm{s}^{-1}$ and with $y$ the distance from centre of the channel in the latitudinal direction. These values are typical of the Southern Ocean but the gradient in planetary vorticity has a smaller contribution to the dynamics than the effective beta induced by topography. Momentum is removed from the bottom isopycnal layer via quadratic drag with a drag coefficient of $c_{\rm drag} =  0.003$.

The idealised channel model configuration is informed by the neutral density profiles and zonal velocities of sections of the ASC from global ocean-sea ice model simulations using the ACCESS-OM2-01 global ocean-sea ice model and from previous idealised experiments \citep{Stewart2015Eddy-mediatedBreak, Stewart2016EddySlope, Huneke2019DeepFront, Kiss2020ACCESS-OM2Resolutions}. We use four density layers to represent the slope current region: two layers of Antarctic Surface Water, a Circumpolar Deep Water (CDW) layer and a Dense Shelf Water layer. The density values used in each of the layers are $\rho_1 = 1027.8 \,\mathrm{kg\, m^{-3}}$, $\rho_2=1028.0 \,\mathrm{kg\, m^{-3}}$, $\rho_3=1028.1 \,\mathrm{kg\, m^{-3}}$, $\rho_4=1028.3 \,\mathrm{kg \, m^{-3}}$, and were based on neutral density values from \citet{Stewart2015Eddy-mediatedBreak}. The model domain is shown in Figure~\ref{fig:model_setup} (a), with a snapshot of the top surface water isopycnal layer, CDW layer, and dense shelf water layer. There are no tides or other water mass transformations. The idealised isopycnal channel model is not able to model water mass transformations, such as the dense shelf water formation in regions characterized by a dense shelf regime, or basal melt in warm shelf regions in both West and East Antarctica. However, there are no water mass transformations in the fresh shelf regime if warm CDW has not yet intruded onto the continental shelf \citep{BoeiraDias2023SensitivityResolution}, enabling us to use this isopycnal channel framework to study episodic canyon intrusions in a fresh shelf regime. Density interfaces are restored at the northern boundary to mimic the open ocean, with the depth of density interfaces at the northern boundary being prescribed in the initial conditions. The initial conditions are motivated by the fresh shelf regime modelled in the work by \citet{Stewart2015Eddy-mediatedBreak}. The density interfaces throughout the rest of the domain are allowed to vary. The constant wind forcing spins up a current which subsequently becomes unstable and gives rise to eddies. The experiments are run until statistical equilibrium is reached. The initial stratification (before spin-up) has a first Rossby deformation radius of 5$\,\mathrm{km}$ on the shelf and 11$\,\mathrm{km}$ offshore, thus eddies are comfortably resolved by our simulations which have a 1$\,\mathrm{km}$ lateral resolution.  We experimented with a 2$\,\mathrm{km}$ resolution configuration but it did not fully resolve the eddying behaviour in the ASC. A preliminary test on doubling the resolution to 0.5$\,\mathrm{km}$ showed little qualitative change in ASC transport and CDW intrusions.

The wind stress forcing is steady and zonally symmetric, with only zonal wind forcing. We use a maximum wind stress input of $\tau_0 = 0.1 \, \mathrm{N}\, \mathrm{m}^{-2}$ for the control simulation, as in the idealised configuration representing the fresh shelf regime in \citet{Stewart2015Eddy-mediatedBreak}. The zonal wind stress, $\tau_x$, is 
\begin{equation}\label{eqn:wind_input_function}
    \tau_x = -\tau_0 \cos^{2}(\pi y / \sigma_\tau) ,
\end{equation}
where $\sigma_\tau = 500\,\mathrm{km}$ is the width of the channel, also the width of the wind stress forcing profile. The zonal wind profile for the control simulation is shown in Figure~\ref{fig:model_setup} (b). There is no seasonal variation in either the surface wind or northern boundary restoring forcing, and no sea ice model is incorporated in this channel configuration of MOM6.

 \begin{figure*}
    \includegraphics[width = \textwidth]{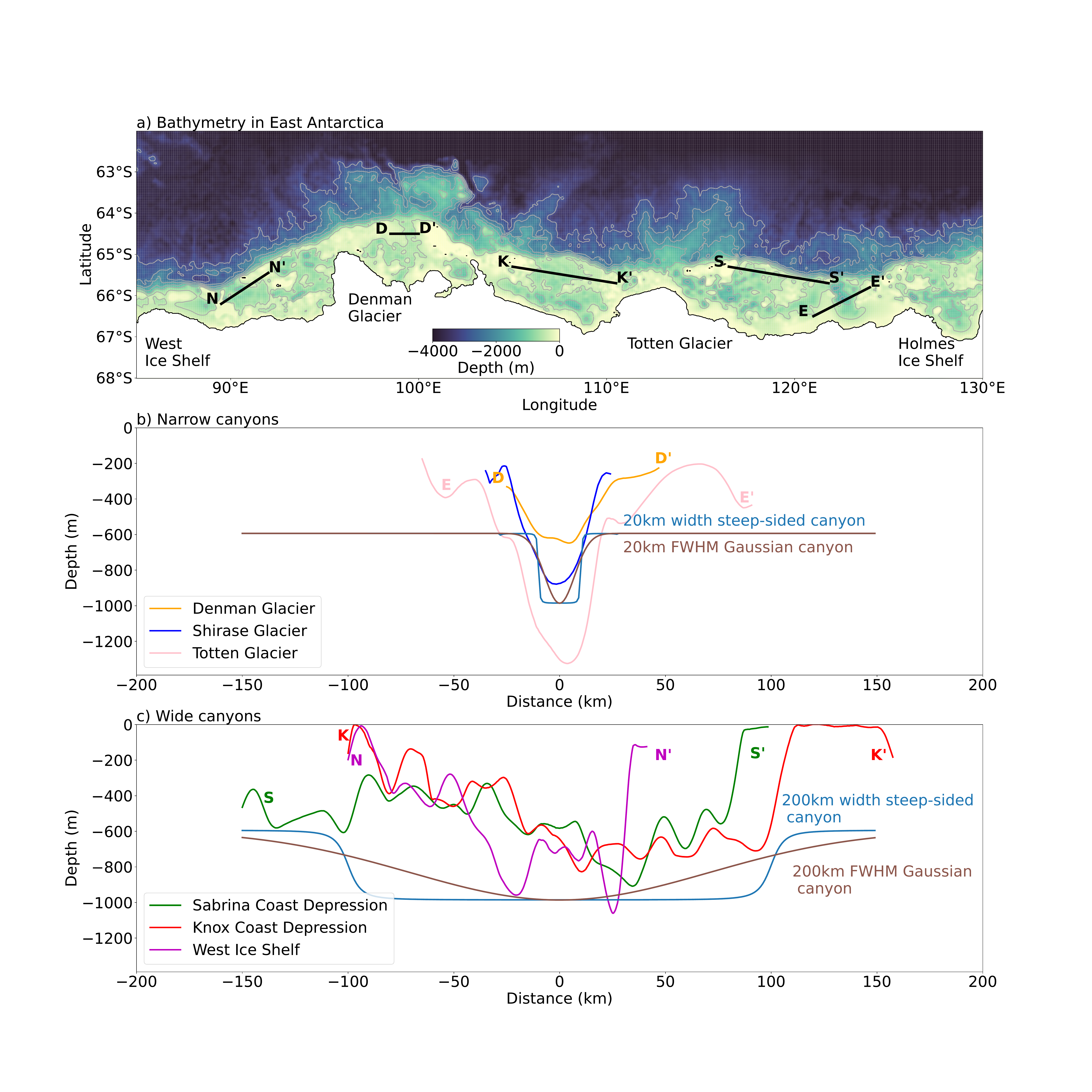}
    \vspace{-4em}
    \caption{(a) Observed bathymetry in the East Antarctic, using ETOPO1 data \citep{NOAANationalGeophysicalDataCenter2009ETOPO1Model}, with sampled canyon cross-sections marked in black. Contours of 500m, 1000m, 1500m and 2000m depth are marked in grey. (b) Cross-sections of observed narrow canyons taken from (a), labelled by  name or associated glacier. The Shirase glacier canyon at 38-39$\degree$ E is not shown in (a) as it is west of the domain pictured. Cross-sections of idealised narrow canyons are shown in the steep-sided (blue) and Gaussian (brown) shapes. (c) Cross-sections of observed wide canyons taken from (a), labelled by name or associated glacier. Cross-sections of idealised wide canyons are shown in the steep-sided (blue) and Gaussian (brown) shapes. }
     \label{fig:observed_canyons}
 \end{figure*}

We add canyons to the continental slope topography to investigate how their presence influences shoreward transport of CDW onto the continental shelf (see Figure~\ref{fig:observed_canyons}). The canyon geometries in this investigation are idealised: we use a steep-sided canyon, shaped like a trough with a relatively flat-bottom in the canyon, or a canyon with sloped-sides of a Gaussian-shape (see blue and brown lines respectively in Figure~\ref{fig:observed_canyons} (b) and (c)). Figure~\ref{fig:model_setup}~(a) shows the topography for a steep-sided canyon of 200$\,\mathrm{km}$ width. Simulations for the steep-sided and Gaussian canyon cases were run using canyons of width 20$\,\mathrm{km}$, 50$\,\mathrm{km}$, 100$\,\mathrm{km}$, 150$\,\mathrm{km}$ and 200$\,\mathrm{km}$ at half of the canyon's maximum depth (analogous to the full width half maximum), and with a canyon depth of 400$\,\mathrm{m}$ relative to the continental shelf. The widths of the canyons are comparable to cross-sectional widths of canyons observed around the Antarctic Margin, as shown in Figure~\ref{fig:observed_canyons} and were chosen from the ETOPO1 Global Relief Model \citep{NOAANationalGeophysicalDataCenter2009ETOPO1Model}. The topography of the continental slope is based on previous idealised experiments and the ACCESS-OM2-01 model \citep[e.g.][]{Stewart2015Eddy-mediatedBreak, Kiss2020ACCESS-OM2Resolutions}).

After channel model spin-up and equilibration, a snapshot of the stratification of the idealised fresh shelf regime is shown in Figure~\ref{fig:model_setup}~(a), for the control simulation with the steep-sided canyon of 200$\,\mathrm{km}$ width, the widest canyon modelled. This stratification is comparable to observations, such as the section of the Eastern Weddell Sea in Figure 3 (d) of \citet{Thompson2018TheClimate}, who used data from \citet{Heywood2002WaterOceans}. Statistical equilibrium was typically reached after 20 years of spin-up, with daily mean data for analysis taken from at least a 20 year long period of statistical equilibrium; each experiment is spun up separately. A strong westward current is spun up in the surface layers with a maximum zonal velocity of 0.3 $\mathrm{m}\,\mathrm{s}^{-1}$ in the water column, when averaged over 10 years, shown in Figure~\ref{fig:model_setup}~(c). ASC velocities in this simulation are slightly higher than velocities in other models and observations. For example, \citet{Huneke2022SpatialModel} showed a maximum zonal velocity of 0.2 $\mathrm{m}\,\mathrm{s}^{-1}$ in the water column in a region of the fresh shelf regime, when averaged over 10 years, while a regional model of the ASC showed maximum instantaneous zonal velocities off the Totten Ice Shelf in East Antarctica to be 0.2 $\mathrm{m}\,\mathrm{s}^{-1}$ \citep{Nakayama2021AntarcticGlacier}. In our simulation, density surfaces are steeply sloped and incrop into the continental slope, consistent again with regions in the fresh shelf regime. Although idealizations are made in the  configuration so that eddies can be resolved and a wide parameter space can be explored, this control simulation still exhibits the key features characteristic of the fresh shelf regime and East Antarctica, and is used as a basis for exploring how CDW intrusions access the Antarctic continental shelf through canyons. 

\section{Temporal variability of CDW intrusions and the ASC}
\label{sec:CDW_intrusions}

 \begin{figure*}
    \includegraphics[width = \textwidth]{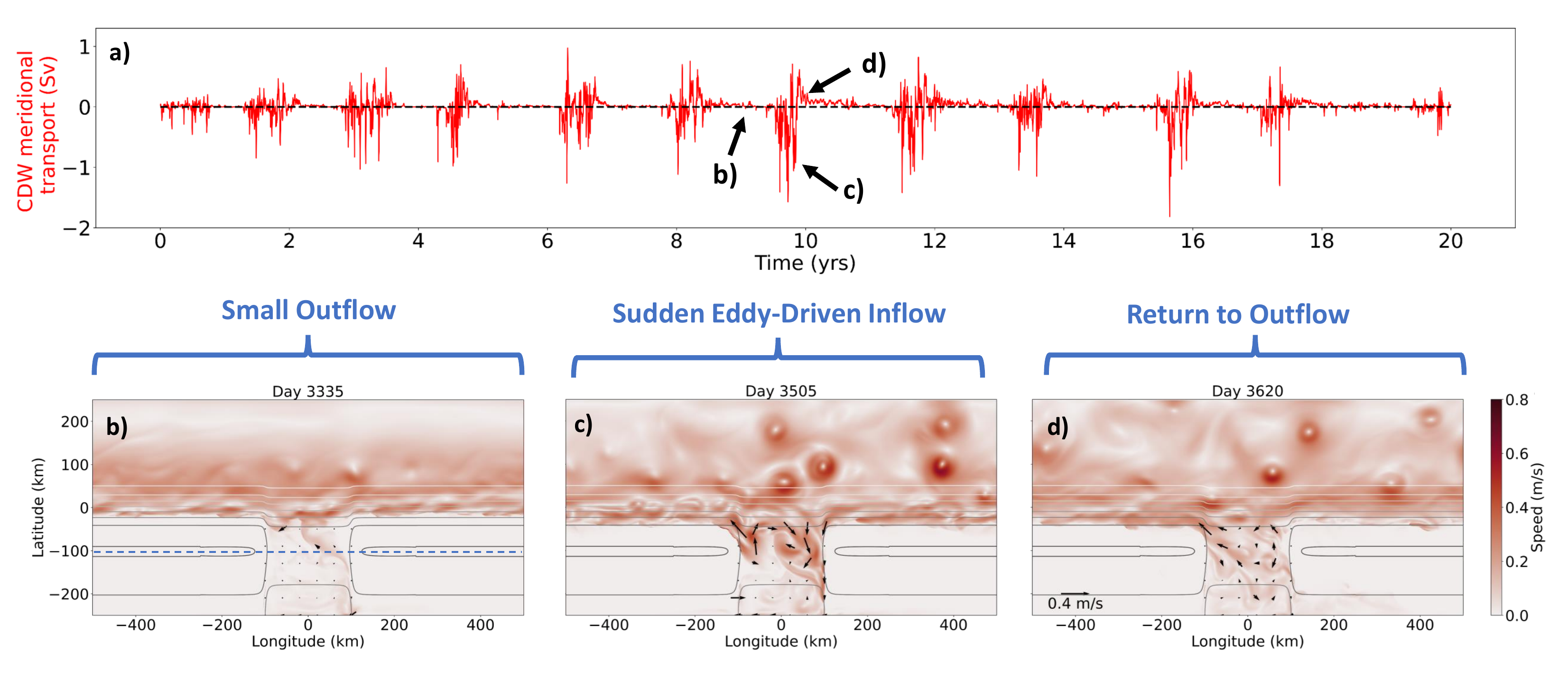}
    \caption{(a) Meridional transport of CDW across a cross-section of canyon at latitude $y = - 100\,\mathrm{km}$ in the control simulation over 20 years at equilibrium. The cross-section is taken across the blue dashed line in (b).  Snapshots of speed in the CDW layer of simulation showing representative examples of the different stages of episodic CDW intrusion: (b) a small outflow of CDW flowing off the continental shelf, (c) a sudden eddy-driven inflow onto the continental shelf, and (d) a return to a CDW outflow. We see that eddies form in the ASC and drive a large onshore flow approximately every two years. The eddies on the continental shelf then begin to dissipate and flow offshore, continuing on the cycle of CDW intrusions. A video showing the episodic CDW intrusions in this experiment is in included in the Supplementary Material.}
     \label{fig:intrusions}
 \end{figure*}

We start by looking at the qualitative behaviour of on-shelf intrusions in the control isopycnal channel simulation of Figure~\ref{fig:model_setup}. We choose the control simulation as this is the canyon configuration which allows for the largest CDW intrusions, and it typifies canyon geometry at certain locations around the East Antarctic sector (Figure~\ref{fig:observed_canyons}~(c)). We are primarily interested in meridional transport of CDW, hence the time series in Figure~\ref{fig:intrusions}~(a) shows meridional transport in the CDW layer at the sill latitude across a section of the canyon (see dotted line in Figure~\ref{fig:intrusions}~(b)), with poleward transport shown as negative. Since there is no water mass transformation on the shelf, the CDW shelf exchange is balanced over a long time period. However, this time series shows that there is poleward meridional transport of CDW in the canyon, and that this cross-shelf transport is not constant but instead occurs in isolated episodes approximately every two years. The different stages of CDW intrusion are highlighted in snapshots of speed in the CDW layer over time, shown in separate stages of Figure~\ref{fig:intrusions} (b), (c) and (d). The most common state of the system is the small outflow state of little cross-shelf exchange as indicated in Figure~\ref{fig:intrusions} (b), with a minimal amount of CDW draining off the continental shelf. There is then a transition to a stronger eddy field in the ASC, corresponding with a stage of sudden eddy-driven flow of CDW onshore, with coherent vortices reaching the continental shelf as seen in Figure~\ref{fig:intrusions} (c). While CDW pools on the continental shelf, eddies in the ASC weaken, and CDW begins to flow offshore, returning to the original state of primarily offshore transport of CDW as indicated in Figure~\ref{fig:intrusions} (d). This entire cycle of weak CDW offshore flow, followed by a sudden shoreward flow, and a return to an offshore flow is surprising given that the simulated ASC has reached a statistically steady state. Additionally, as wind and sponging of layer interfaces remain constant throughout the simulation, this interannual variability of episodic CDW intrusions is not the result of external forcings, pointing to the presence of intrinsic temporal variability. Previous studies have all focused on the initial driver of CDW intrusions: \citet{St-Laurent2013OnShelves} for example, showed that wave-topographic interactions allowed for a steady anti-cyclonic circulation in the canyon to develop in an idealised model of the Marguerite Trough of the Bellingshausen Sea. However, no previous work has pointed to mechanisms driving an intrinsic variability in CDW intrusions independent of external forcings. Understanding this CDW transport variability is the focus of the rest of this section, initially analysing the effect of canyon width on the variability of CDW intrusions, then investigating the origins of the CDW variability.

 \begin{figure*}
    \centering
    \includegraphics[width = \textwidth]{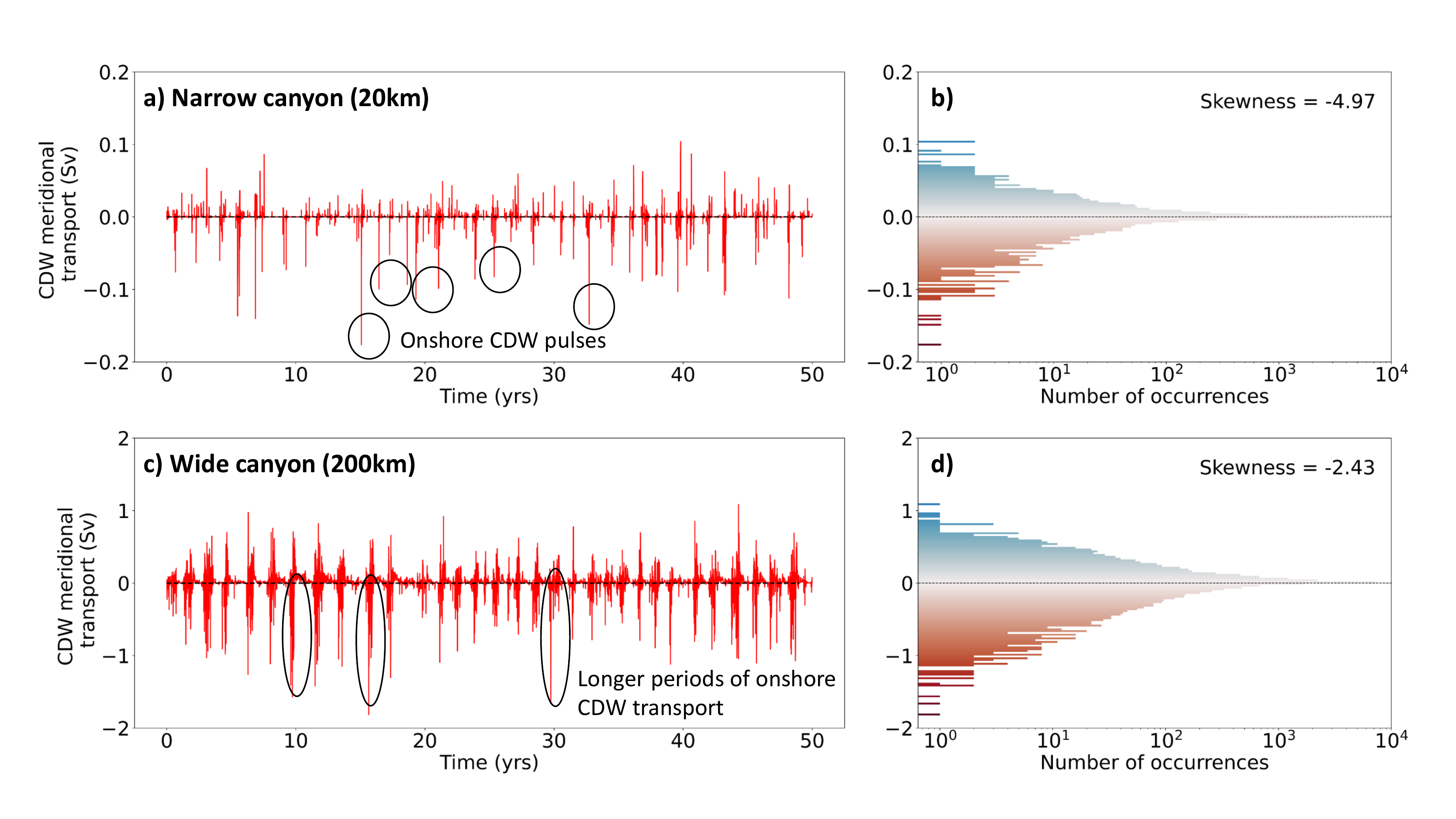}

    \caption{(a) Meridional transport of CDW across latitude $y = - 100\,\mathrm{km}$ and (b) probability density function of meridional transport of CDW across the same latitude for the narrow 20$\,\mathrm{km}$ wide steep-sided canyon case. (c) and (d) show the same quantities as (a) and (b) respectively but for the wide 200$\,\mathrm{km}$ steep-sided canyon case. Skewness values of meridional CDW transport indicate that a more negatively skewed distribution has more irregular and episodic CDW intrusions. The peaks circled show examples of isolated pulses of CDW onshore in (a) and longer periods of intrusions in (c). The colors in the probability distribution function of (b) and (d) qualitatively represent CDW flowing onshore and heating the shelf in red, and CDW leaving the shelf in blue.}
    \label{fig:histogram}
\end{figure*}

\subsection{Effect of canyon width on warm water intrusions}
We observe in the control simulation the intrinsically episodic behaviour of CDW intrusions through a canyon, and see in Figure~\ref{fig:intrusions} that the canyon is a key pathway by which CDW reaches the continental shelf. Hence we aim to understand how characteristics of the canyon, specifically the width of the canyon, affect the episodic poleward flow of CDW. We first compare the narrowest (20$\,\mathrm{km}$ width) and widest (200$\,\mathrm{km}$ width) steep-sided canyon cases of the simulations we have run. The time series of meridional CDW transport at the sill latitude is plotted in Figure~\ref{fig:histogram}, with the time series for the narrow canyon case in (a) and that of the widest canyon case in (c). Qualitatively, the narrower canyon case shows more isolated instances of CDW intrusions highlighted by circles in Figure~\ref{fig:histogram} (a). However, CDW intrusions in the wider canyon case last for a longer period of time, instead of the isolated single bursts of CDW travelling onto the shelf in the narrow canyon case, shown in the ovals on Figure~\ref{fig:histogram} (c). A plot of the probability density function for meridional CDW transport using 50 years of daily data reveals a isolated pulses of large poleward transports in the narrow canyon case (Figure~\ref{fig:histogram} (b) for the narrow 20$\,\mathrm{km}$ and (d) wide 200$\,\mathrm{km}$ case). The isolated pulses of CDW in the narrow canyon case are separated from the central distribution, causing the distribution to be heavily skewed. Hence, a topographic configuration with more asymmetric, or irregular, CDW intrusions has a more negatively skewed distribution of meridional CDW transport.

\begin{figure}
    \centering
    \includegraphics[width = \columnwidth]{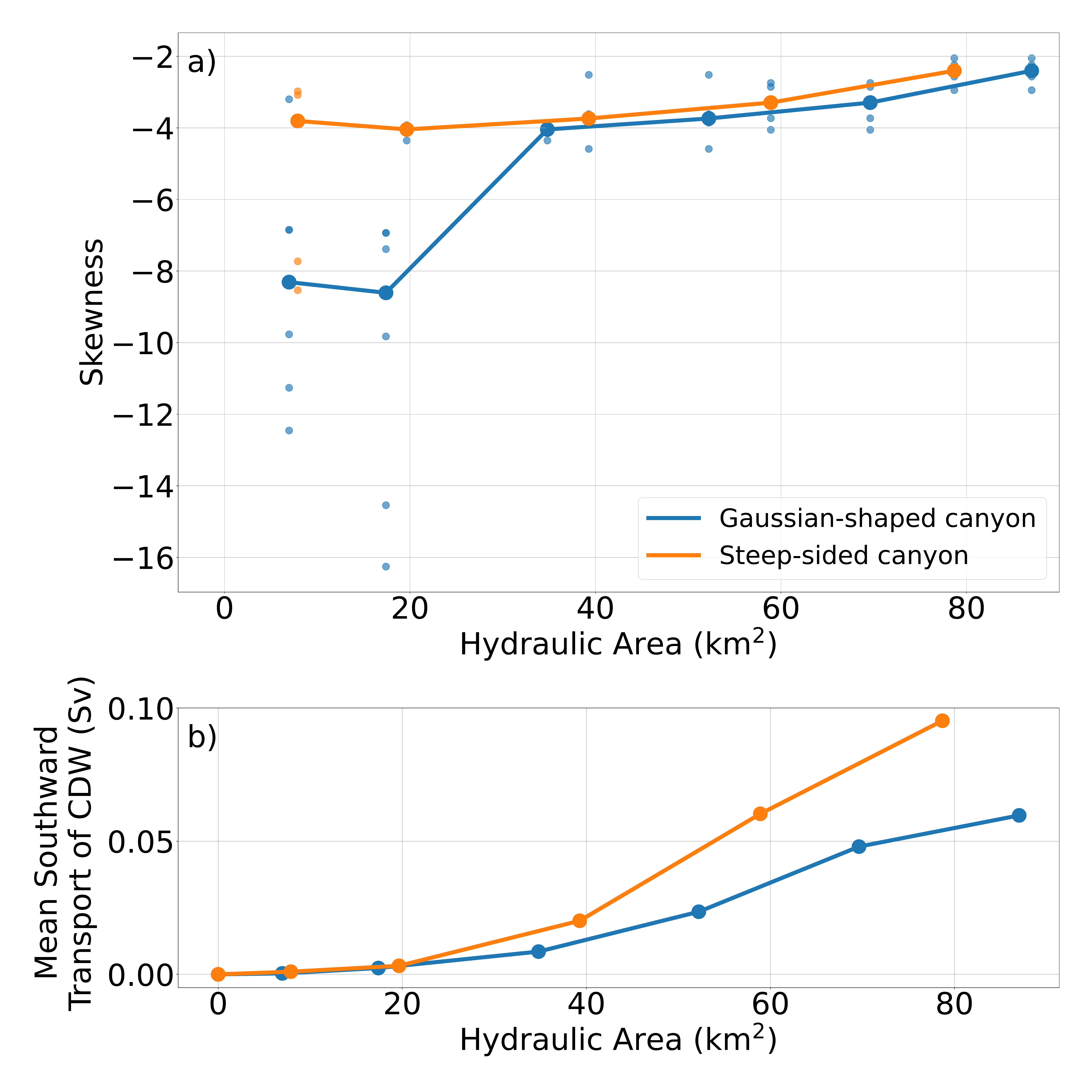}
    \caption{(a) Skewness of meridional CDW transport distribution as in Figure~\ref{fig:histogram} with hydraulic area of canyon, for canyons of steep-sided and Gaussian-shapes. Skewness calculations for each canyon configuration use subsampled sections of 20 year length, overlapping by 10 years. The low opacity, smaller points are skewness values for subsampled sections, while the median skewness values for each canyon configuration are plotted in the solid color. Narrower canyons have a more negatively skewed distribution, with greater asymmetry and irregularity in CDW transport. For the two smallest Gaussian canyons and the narrowest steep-sided canyon cases, we run experiments for 70 years instead of 50 years, to improve the statistics in narrow canyons with irregular onshore flow. (b) Mean southward transport of CDW across latitude $y =-100\,\mathrm{km}$ with hydraulic area of canyon for canyons of steep-sided and Gaussian-shapes.  Canyons with a larger hydraulic area allow for a greater transport of CDW poleward and more regular intrusions.  }
    \label{fig:poleward-transport}
\end{figure}

To directly compare the regularity of CDW transport across different canyon widths, we use the Fisher-Pearson coefficient of skewness to quantify the asymmetry of CDW flow on and off the shelf. The results of this comparison are shown in Figure~\ref{fig:poleward-transport} (a), plotting the skewness, analogous to the regularity of intrusion, against the hydraulic area of the canyon. Skewness calculations at each canyon configuration use four subsampled sections of 20-year long time-series, overlapping by 10 years, from which a median skewness is calculated. For the narrowest canyons, a few isolated intrusion events cause a large spread in skewness. Therefore to increase the robustness of the skewness metric, we run the smallest two Gaussian canyons and the narrowest steep-sided canyon cases longer for a 70-year long statistically equilibrated timeseries, improving the skewness statistics with two more subsampled sections. We use skewness as a metric because simulations with smaller canyons have increasingly rare onshore CDW transport events, reflected in a more negatively skewed distribution. We have defined the hydraulic area to be the cross-sectional area of the canyon at the shelf break, at the latitude $y = - 100\,\mathrm{km}$. For the steep-sided canyon experiments (orange line), we find that narrower canyons have a more negatively skewed distribution of meridional CDW transport, and thus more asymmetric CDW transport. Narrow steep-sided canyons have sudden large intrusions but more frequent instances of small CDW drainage, while wider steep-sided canyons exhibit less asymmetry between the CDW drainage and intrusions onto the shelf. Hence, the frequency of CDW flow onshore is relatively steady in wide canyons, which can improve the predictability of intrusions. Observed canyons do not always have geometries comparable to the steep-sided case, so we conducted the same analysis for experiments with canyons of a more gently-sloped Gaussian-shape with varying width, plotted in blue in both panels of Figure~\ref{fig:poleward-transport}, where we confirm that the same overall trend in episodic behaviour with canyon width holds as in the steep-sided canyon cases. Note that additional experiments run with wind forcing shifted north/south by 50$\textrm{km}$ reveals that the skewness is more sensitive to canyon geometry than modest shifts in ASC position.  

Alongside understanding the nature of CDW intrusions and its variability, we also compare poleward CDW transport between different canyon configurations. We select the southward transport values of CDW at the sill latitude, zero out any northward transport, and zonally-integrate the southward transport at each time step before taking the temporal mean over 50 years of daily data. The time-mean southward transport of CDW is computed in simulations with a range of canyon widths and canyon geometries, including a simulation without a canyon on the continental slope, as seen in Figure~\ref{fig:poleward-transport} (b). Across both canyon geometries, southward transport of CDW onto the shelf increases with the hydraulic area available for flow onto the shelf. Note that in the simulation without a canyon, there is no southward transport of CDW across the sill latitude, highlighting the importance of canyons in CDW intrusions. Additionally, canyon configurations with steep-sided canyons have a greater southward transport of CDW than those with a Gaussian-shape shown in blue in Figure~\ref{fig:poleward-transport}~(b), even when canyons with the same hydraulic area are compared. We conclude that wider canyons allow for more CDW transport onto the shelf, with less asymmetry in meridional CDW transport when compared to narrower canyons, and this result is robust for the two canyon geometries tested. 

The effect of canyon width on the regularity of CDW intrusions can affect the predictability of the East Antarctic Margin: although narrow canyons allow for less CDW transport poleward, the asymmetry towards intrusions and the key locations of narrow canyons, e.g. beneath the Denman Glacier, may indicate that further study into cross-shelf dynamics across narrow canyons is required in order to predict the variability of warm water intrusions onto the shelf where vulnerable ice shelves sit. Wider canyons allow for greater CDW transport, with intrusions occurring periodically at a set frequency, which may make flow at these wider canyons easier to predict. We next investigate the mechanisms governing the periodic behaviour of CDW intrusions. 

\subsection{Link between CDW intrusions and ASC variability}

 \begin{figure*}
    \centering
    \includegraphics[width = 27pc]{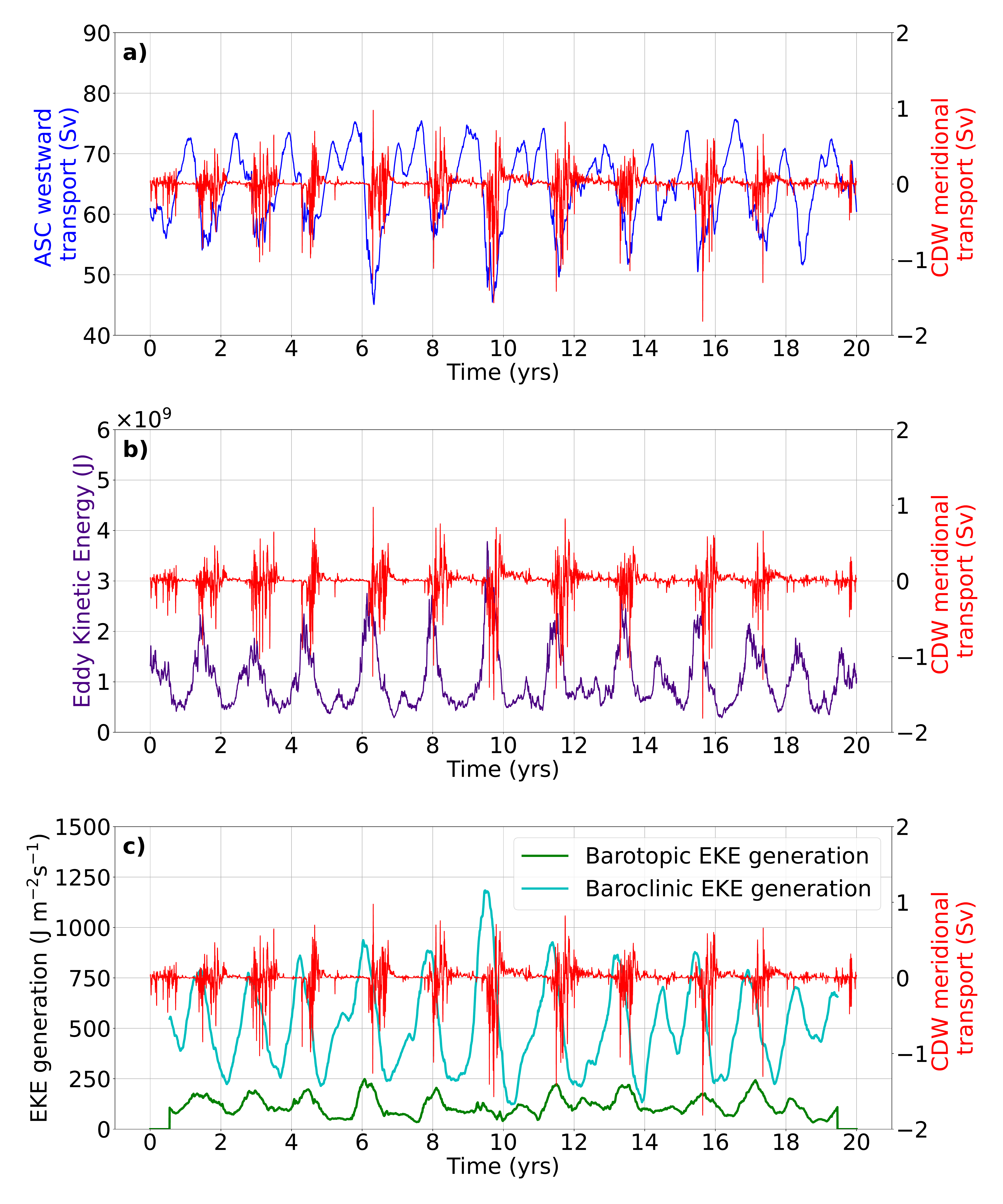}
    \caption{(a) Time series of ASC strength and meridional volume transport of CDW onto shelf at latitude $y =- 100\, \mathrm{km}$. (b) Time series of EKE per unit area and the same CDW transport time series as in (a). (c) Time series of baroclinic and barotropic energy conversions per unit area and the same CDW transport time series as in (a). Energy conversions are calculated as a rolling average of 200 days. All time series are for 20 years after the flow has equilibrated. The episodes of meridional CDW transport coincide with weakening of the ASC, and precedes peaks in EKE and EKE generation by a short time lag of days.}
    \label{fig:time_series}
\end{figure*}

Although we have found an intrinsic temporal variability of CDW intrusions across a range of canyon geometries, the mechanisms governing the variability are unclear. We can nonetheless see a connection between the variability of CDW intrusions and the ASC strength in Figure~\ref{fig:time_series}~(a), where the strength of the ASC is plotted in blue and the meridional CDW transport in red, both plotted for 20 years at statistical equilibrium. In the control simulation with the widest canyon, the ASC strength varies from 40-80 Sv with a mean of around 60 Sv, larger than the ASC strengths observed by \citet{Pena-Molino2016Direct113E}, which varies between 0-100 Sv of westward transport with a mean of around 20 Sv. The ASC strength exhibits variability on an interannual time-scale, and in this control simulation, the poleward transport of CDW is maximised when the slope current is weaker. 

The simulated interannual variability in the ASC is unlike those observed in existing studies. Regional simulations of the Totten Glacier region with time-varying external wind forcings have previously shown a temporal variability in ASC strength, with strong ASC weakening events occurring about every 10 years, but also a higher frequency variability on the order of years \citep{Nakayama2021AntarcticGlacier}. Direct observations of the ASC in East Antarctica have instead primarily found higher-frequency variability in ASC strength \citep{Pena-Molino2016Direct113E}, however, these observed timeseries are shorter than two years, which would not be sufficient to capture biennial oscillations in ASC strength. However, an increased poleward transport of CDW under a weakened ASC is consistent with the results by \citet{Nakayama2021AntarcticGlacier}, and posits a link between episodic warm water intrusions and the ASC strength. 

Most relevant to fresh shelf regime regions around East Antarctica, intrusions of warm water onto the Antarctic continental shelf have previously been found to be driven by eddy activity \citep{Nst2011EddySea,St-Laurent2013OnShelves,Hattermann2014Eddy-resolvingOcean, Stewart2015Eddy-mediatedBreak,Stewart2016EddySlope}. To determine if the CDW intrusions arise from eddies in the ASC, we compute the area-integrated eddy kinetic energy (EKE) in the ASC in the CDW layer. We use a layered thickness-weighted framework \citep{Young2012AnEquations}, with energy transfers between eddy energy reservoirs as described by \citet{Aiki2016EnergeticsEddies} and \citet{Yung2022TopographicUpwelling}. The time-mean EKE in $i$-th~layer is $\frac{1}{2} \rho_0 \overline{h_i|\mathbf{u}''_i|^2}$, where $\rho_0$ is the reference density and we use the density of the top layer as the reference density, $h_i$ is the thickness of layer $i$, $\mathbf{u}_i$ is the velocity in layer $i$, and $\mathbf{u}''_i = \mathbf{u}_i - \widehat{\mathbf{u}_i}^{t}$ is the deviation from the thickness-weighted mean velocity $\widehat{\mathbf{u}_i}^{t}$, where $\widehat{\mathbf{u}_i}^{t} = \overline{h_i \mathbf{u}_i}^{t} / \overline{h_i}^{t}$, with $\overline{(\cdot)}^{t}$ a time-mean. The EKE is integrated over the ASC in the CDW layer, bounded by the latitudes of $y = - 50 \, \mathrm{km}$ and $y = 100 \, \mathrm{km}$ and integrated across the whole channel in the zonal direction. Inspecting the time series of the EKE with the meridional volume transport of CDW in Figure~\ref{fig:time_series} (b), plotted for 20 years at equilibrium, we observe that episodes of southward transport of CDW follow peaks in area-integrated EKE. The increased southward transport of CDW thus appears to be linked to a greater presence of eddies in the ASC, such that the temporal variability in rates of eddy generation could drive the intrusions of CDW. 

To identify the mechanisms driving the temporal variability of EKE, we compute the baroclinic and barotropic contributions of energy conversion to EKE in the ASC \citep{Aiki2016EnergeticsEddies, Yung2022TopographicUpwelling}. The baroclinic energy conversion term, $\overline{\mathbf{u}'_i \cdot ( h_i \boldsymbol{\nabla} \phi'_i)}$,  links the contribution of interfacial form stress to EKE in layer~$i$, while the barotropic energy conversion term, $\rho_0 (\widehat{\mathbf{u}_i} \cdot \boldsymbol{\nabla}) \cdot (\overline{h_i \mathbf{u}'_i \otimes \mathbf{u}'_i})$, is the contribution of Reynolds stress, and thus horizontal shear, to EKE. Here, $\phi_i$ is the Montgomery potential and $\phi'_i = \phi_i - \overline{\phi_i}$, where $\overline{(\cdot)}$ is a rolling mean over 200 days to smooth out transient eddy effects. $\mathbf{u}'_i = \mathbf{u}_i - \widehat{\mathbf{u}_i}$ is the thickness weighted mean velocity computed using a rolling mean, where $\widehat{\mathbf{u}_i} = \overline{h_i \mathbf{u}_i} / \overline{h_i}$, and  $\otimes$ is the outer product of two vectors.

The baroclinic and barotropic energy conversions for the control simulation of the widest canyon are plotted in Figure~\ref{fig:time_series}~(c) for 20 years at equilibrium. The peaks in both the baroclinic and barotropic energy conversions precede peaks in warm intrusions, showing that the generation of eddies results in more CDW intrusions with a short time lag of days. Crucially, the baroclinic energy conversion term dominates the barotopic energy conversion term, highlighting the contribution of baroclinic instability to the EKE gain. Hence, this baroclinic instability is the mechanism by which eddies are generated to drive CDW intrusions on to the shelf.

We conclude that episodic generation of eddies through baroclinic instability weakens the ASC, while simultaneously driving the onshore transport of CDW. Previous work by \citet{Nakayama2021AntarcticGlacier} has already shown a link between a weak ASC and warm CDW intrusions, and in this channel model, a weaker ASC is linked with warm CDW intrusions by the generation of eddies through baroclinic instability. The mechanism is further supported by experiments with increased bottom drag, under which eddy generation is reduced and CDW intrusions are simultaneously inhibited (not shown). As the variability of CDW intrusions originates from changes in the generation of eddies in the ASC, the ASC is therefore a key governing factor of CDW intrusions whose intrinsic variability is crucial to be understood.

\section{Intrinsic ASC variability}
\label{sec:ASC_variability}
In Section~\ref{sec:CDW_intrusions} we demonstrated that there is an intrinsic time-variability to CDW intrusions caused by a cycle of rising and falling rates of eddy generation in the ASC. However we did not address the origin of the variability in the ASC. In this section, we show that the cycle of rates of eddy generation in the ASC can be explained using a low-order model, consisting of two non-linear ordinary differential equations, that describes the coupling of available potential energy and eddy kinetic energy, and that the model predicts an intrinsically oscillatory ASC when linearised. We also compare the low-order model predictions to our isopycnal channel model simulations, firstly by comparing the energy evolution predicted by the low-order model to channel model simulations, followed by parameter sensitivity tests. These comparisons allow us to evaluate whether the physics of the low-order model can explain the key processes governing ASC time-variability in the fresh-shelf regime.

\subsection{Low-order model and oscillatory solution to linearised equations}
We demonstrated that the intrinsic variability of the idealised ASC in the fresh shelf regime arises from eddy generation via baroclinic instability. However, the rate of baroclinic eddy generation is not constant even at equilibrium, instead, baroclinic instability occurs in a periodic cycle. The cause of this cycle and its timescale are unclear as there are no time-varying external forcings to induce this oscillatory behaviour. Therefore, we aim to understand the equilibrated state better, prompting the use of a low-order model. 

\begin{figure}
    \centering
    \includegraphics[width = 19pc]{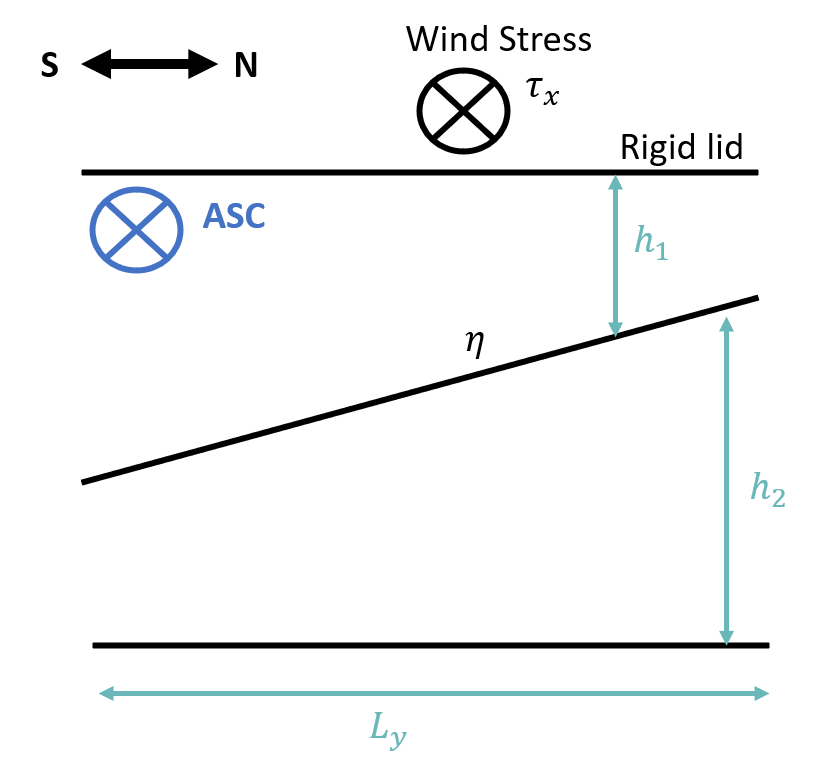}
    \caption{Schematic showing features of the low-order model, replicating an ASC at equilibrium, with two density layers, a rigid lid, a sloping isopycnal of interface height $\eta$ with a constant slope across the extent of the domain $L_y$, zonal wind stress input of $\tau_x$, layer thicknesses of each layer being $h_1$ and $h_2$ for top and bottom layers respectively. The wind input slopes the isopycnal interface and spins up a westward current.} 
    \label{fig:2layer_energy_ODE_setup}
\end{figure}

We postulate that the interannual variability in this system occurs due to energy exchange between total eddy energy and total available potential energy. To demonstrate this, we develop a low-order system that governs the evolution of total eddy energy, and total available potential energy in an idealised ASC, in order to understand the quasi-steady state of the ASC. We start with a two-layer, zonally symmetric isopycnal model of the ASC, pictured in Figure~\ref{fig:2layer_energy_ODE_setup}, with constant zonal wind forcing. The fluid interface is assumed to have a constant slope. The energy budget of the system can be expressed as: 

\begin{align}
    \frac{\mathrm{d}E}{\mathrm{d}t} &= \textrm{eddy energy conversion} - \textrm{damping}, \label{eqn:word_Eeqn} \\
    \frac{\mathrm{d}APE}{\mathrm{d}t} &= \textrm{wind input} - \textrm{eddy energy conversion}, \label{eqn:word_APEeqn}
\end{align}

\noindent with

\begin{align}
    E &= \sum_i \int \large( \tfrac1{2} \rho_0 h_i |\mathbf{u}''_i|^2 +  \tfrac1{2} \rho_0 g' \eta''^2 \large)    \mathrm{d} y, \label{eqn:Edef}\\ 
    APE &= \sum_i \int \tfrac1{2} \rho_0 g' \eta^2 \mathrm{d} y,  \label{eqn:APEdef}
\end{align}
where $E$ is the depth-integrated total eddy energy per unit length in the longitudinal direction and $APE$ is the depth-integrated total available potential energy per unit longitudinal length. $\eta$, the isopycnal interface height, is defined relative to the initial interface height at the southern boundary and $\eta'' = \eta - \overline{\eta}^{t}$. Damping is in the form of $\lambda E$, with $\lambda $ the linear damping coefficient. Eddy energy conversion is $ \vert f \vert/N \, \vert\partial u/\partial z\vert \, E$ (following \citet{Marshall2017EddyCurrent}), where $f$ is the Coriolis parameter, $N$ is the buoyancy frequency and $u$ is the zonal velocity. Wind input per unit longitudinal length is given by $\int \boldsymbol{\tau} \cdot \mathbf{u}_{\rm surface} \, \mathrm{d}y$, where  $\boldsymbol{\tau}$ is the wind stress vector and $\mathbf{u}_{\rm surface}$ is the surface velocity.

We wish to express~\eqref{eqn:word_Eeqn} and~\eqref{eqn:word_APEeqn} as coupled system of differential equations in terms of $APE$ and $E$, so we define the components of the equation in terms of these variables. We assume geostrophic velocities in each of the isopycnal layers, thermal wind balance, rigid lid, constant linear shear and zero bottom velocity. The zero bottom velocity is justified as bottom velocities in both our channel simulations and in observations around the East Antarctic are generally less than 10$\%$ of the surface velocities.  The procedure for defining and discretising these terms for the case of the two-layer model is outlined in the Appendix, and we end up with the equations: 

\begin{align}
    \frac{\mathrm{d}E}{\mathrm{d}t} &= \underbrace{2\sqrt{\frac{6}{\rho_0 H L^{3}_y}}}_{a} \sqrt{APE} \, E - \lambda E, \label{eqn:E_eqn} \\
    \frac{\mathrm{d}APE}{\mathrm{d}t} &= \underbrace{\frac{2|\tau_x|}{|f|} \sqrt{\frac{6g'}{\rho_0  L_y}}}_{b} \sqrt{APE} - \underbrace{2 \sqrt{\frac{6}{\rho_0 H L^{3}_y}}}_{a} \sqrt{APE}\, E, \label{eqn:APE_eqn}
\end{align}

\noindent where $\rho_0$ is the reference density, $g' = g \Delta \rho / \rho_0$  is reduced gravity of the two layer system, $H$ is the total thickness of fluid, $L_y$ is the latitudinal extent of the domain where the slope of the density surface is constant, and $\tau_x$ is the zonal component of the wind stress input. We also use that for a two-layer model $N^2 = 2g'/H$. 
 
Now we have two coupled non-linear ordinary differential equations~\eqref{eqn:E_eqn} and~\eqref{eqn:APE_eqn}. To find the fixed points we set $\mathrm{d}/\mathrm{d}t = 0$ and solve, setting the constants to $a$ and $b$ for simplicity. We get a non-trivial fixed point:

\begin{equation}
\sqrt{APE} = \lambda/a \quad \textrm{and} \quad E = b/a . \label{eqn:fixedpoint}
\end{equation}

Upon linearising the energy equations~\eqref{eqn:E_eqn}-\eqref{eqn:APE_eqn} about the equilibrium~\eqref{eqn:fixedpoint}, we obtain evolution equations for the deviations of $E$ and $APE$ about their equilibrium values. Combining these evolution equations we end up with:
\begin{equation}
    \frac{\mathrm{d}^2E}{\mathrm{d}t^2} = - \frac{ab}{2} E,  
\end{equation}

\noindent which has oscillatory solutions of the form $E \propto e^{i \omega t}$, with 

\begin{align} \label{eqn:period}
    \omega & = \sqrt{\frac{ab}{2}} = \frac{1}{L_y}\sqrt{\frac{6 |\tau_x| }{|f| \rho_0 }}\left(\frac{2g'}{H}\right)^{1/4},\\
    \mathrm{T} &= \frac{2 \pi}{\omega} = \pi L_y \sqrt{\frac{2\rho_0 |f|}{3 |\tau_x| }}\left(\frac{H}{2g'}\right)^{1/4},
\end{align}

\noindent where $\omega$ is the radial frequency of oscillation, and $\textrm{T}$ is the period of the oscillation of $E$ and $APE$.

\begin{figure*}
    \centering
    \includegraphics[width = \textwidth]{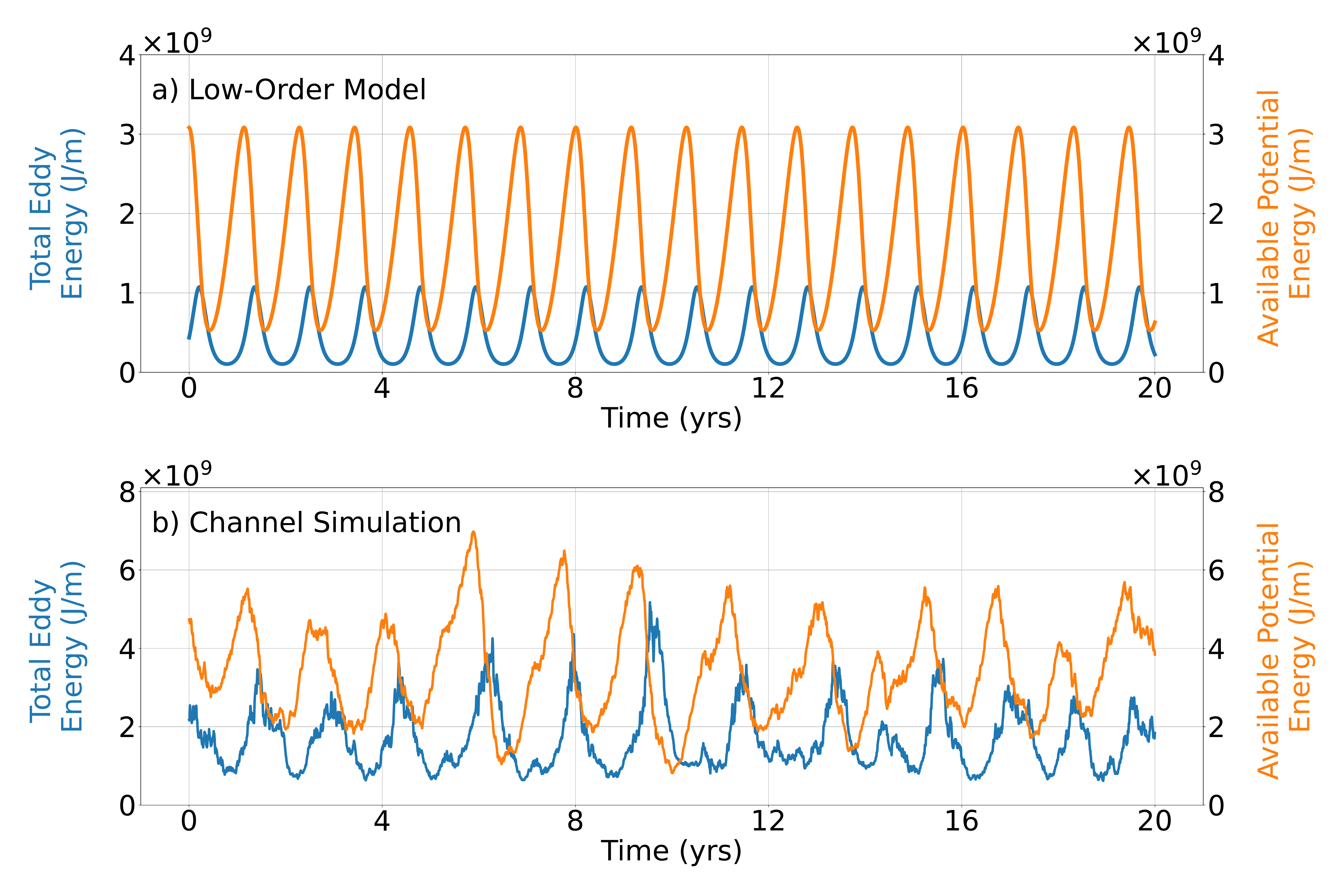}
    \caption{(a)  Solution to initial value problem of the low-order model, using $\lambda = 5 \times 10^{-7} \, \mathrm{s}^{-1}$ and initial conditions for $E$ and $APE$ are 1.05 times and 2 times their equilibrium values in Equation~\eqref{eqn:fixedpoint} respectively. (b) Diagnostics of Eddy Energy and Available Potential Energy, calculated using 20 years of data at equilibrium from the control channel simulation (widest 200km steep-sided canyon). $E$ is calculated in the CDW layer, and $APE$ is calculated at the interface between the CDW layer and the lower surface water layer in the channel model simulations. Although the time-series has a period greater than that predicted around the equilibrium state using the low-order model, the $APE$ similarly peaks and declines before $E$ begins to increase as eddies are being generated.}
    \label{fig:IVP_mom6}
\end{figure*}

The solution to the linearised equations about the fixed point is oscillatory, which is consistent with an intrinsically time-varying ASC. The oscillatory solution implies that the total eddy energy and available potential energy in the ASC will follow an oscillation with a steady period defined in~\eqref{eqn:period}, despite no timescale being imposed on the model through external forcing. The next section will compare the low-order model to the channel model simulations, such that conclusions about the underlying dynamics can be drawn between these system.

\subsection{Energy evolution in low-order model}\label{sec:ivp_verify}

The low-order model~\eqref{eqn:E_eqn}--\eqref{eqn:APE_eqn} describes the evolution of eddy energy and available potential energy in a current system, and when the system is close to equilibrium the low-order model acts as an oscillator. In order to evaluate the extent to which the low order model replicates the physics in the isopycnal channel model simulations, we compare key characteristics of the low-order model to the simulations. We solve the low-order model fully to obtain the time-evolution of energy reservoirs, starting from prescribed initial conditions. The predicted energy evolutions can then be compared to their corresponding quantities in the isopycnal channel model simulations. 

Although the linearised low-order model does show an oscillation consistent with the channel model simulations, there are a number of assumptions in the low-order model that do not hold in the simulation, making direct comparisons difficult. For example, the low-order model assumes a constant slope of the density interface throughout the slope current and a linear drag, while the isopycnal channel simulations have a varying interfacial slope and quadratic drag. Additionally, the simulations feature continental slope topography, which can suppress the generation of eddies \citep{Isachsen2011BaroclinicSimulations}, while the low-order model assumes a flat-bottom channel. The lack of topographic variation in the low-order model implies that all wind stress input has to be balanced by bottom drag, rather than being balanced by topographic form stress in a setup with bathymetric features \citep{Munk1951NoteCurrent}. Hence if both setups were to have similar equilibrated zonal transport values, then the low-order model would require much higher values of drag as it does not have any other way to dissipate zonal momentum. Expecting high values of $\lambda$ in the low-order model, we choose $\lambda = 5 \times  10^{-7}  s^{-1}$ as this gives us energy values comparable to the control simulation in Figure~\ref{fig:IVP_mom6}, and we use this bottom drag parameter to solve the non-linear energy equations of the low-order model. 

Solving the initial value problem for the non-linear energy equations in~\eqref{eqn:E_eqn} and~\eqref{eqn:APE_eqn} using the chosen bottom drag parameter, we find that the theoretical solution for $E$ and $APE$ from the low-order model (Figure~\ref{fig:IVP_mom6}~(a)) is comparable to the $E$ and $APE$ time series calculated from the experimental data (Figure~\ref{fig:IVP_mom6}~(b)). The chosen initial conditions for $E$ and $APE$ are 1.05 times and 2 times their equilibrium values in Equation~\eqref{eqn:fixedpoint} respectively, which are representative of the channel model simulations as the maximum $APE$ in the control simulation is double the equilibrium $APE$. In the low-order model solutions, $APE$ increases before $E$ and begins to drop off once $E$ starts to increase. This is similar behaviour to that seen in the channel model simulations, where $APE$ gain precedes the increase in $E$ as plotted in Figure~\ref{fig:IVP_mom6}~(b), supporting the mechanism that $APE$ is being converted into eddy energy via baroclinic instability. A discrepancy between the solution to the initial value problem and the time series of energies from channel model simulations is the period of the oscillation. Although both plots in Figure~\ref{fig:IVP_mom6} show a cycle of regular oscillations, the period of oscillation in the channel model simulations is longer than that expected from the low-order model. Later in this section we show that the period of oscillation predicted by the low-order model is consistently shorter than that in the channel model simulations, and we also investigate the trends in the period of oscillation as experimental parameters are varied.

\subsection{Parameter sensitivity tests in the low-order model}
Having evaluated the solutions to the low-order model, showing the time-evolution of the eddy energy reservoirs, we now investigate how changing parameters affects the predictions of the low-order model. In particular, we look at how predictions for period and equilibrium energies scale with parameters relevant to a changing ASC, and compare these predictions with the isopycnal channel model simulations.  In the solution to the linearised energy equations~\eqref{eqn:period}, the frequency of the intrinsic oscillation is proportional to $\sqrt{\tau_x}$ but independent of linear bottom drag $\lambda$. Additionally, the equilibrium energies in the low-order model~\eqref{eqn:fixedpoint}, $E_0$ and $APE_0$, are proportional to wind stress input $\tau_x$ and bottom drag squared, $\lambda^{2}$. Testing the parameter sensitivity of the channel model simulation as informed by the low-order model is significantly simpler and allows us to assess how the oscillation in the simulated ASC is affected by changing environmental conditions, hence, the focus on the effect of wind stress and bottom drag. 

We modify wind stress and bottom drag in our isopycnal channel model of the fresh shelf regime to test the dependence of the oscillation period and equilibrium energies on these parameters and verify the energy equations of the low-order model. We conducted four additional simulations for wind stress, at $\tau_0 = 0.025, \; 0.05, \; 0.075, \; 0.125\;\mathrm{N}\,\mathrm{m}^{-2}$, and five for bottom drag, with the quadratic drag coefficient at $c_{\rm drag} = 0.0015,\; 0.0024,\; 0.0036, \; 0.0045, \; 0.006$, corresponding to a halving, 20$\%$ decrease, 20$\%$ increase, 50$\%$ increase and doubling of the control simulation's bottom drag coefficient. All these simulations utilise the topography of the widest canyon in the control simulation shown in Figure~\ref{fig:model_setup}, as it exhibits the most regular periodic oscillation in ASC strength in our experiments and thus is the best comparison to the low-order model.  A previous study by \citet{Stern2015InstabilityBreak} also showed an aperiodic cycle in an idealised jet, modelled on the Antarctic Circumpolar Current on the West Antarctic Peninsula continental shelf. However, we do not consider these simulations without a canyon, as preliminary experiments without a canyon did not reveal any regular oscillatory behavior in ASC strength. Understanding how an oscillation in ASC strength is induced uniquely on a continental slope with a canyon will be the subject of future work. 

\begin{table}[t]
\caption{Experimental parameters used in the low-order model.}\label{t1}
\begin{center}
\begin{tabular}{cc}
\hline\hline
Parameter & Value \\
\hline
 $\tau_x$ & 0.05 $\mathrm{N}\,\mathrm{m}^{-2}$ (domain-averaged of \eqref{eqn:wind_input_function}) \\
 $H$ & 3000 m \\
 $\rho_0$ & 1027.8 $\mathrm{kg}\, \mathrm{m}^{-3}$  \\
 $\Delta\rho$ & 0.5 $\mathrm{kg}\, \mathrm{m}^{-3}$\\
 $L_y$ & 25 $\,\mathrm{km}$  \\
 $h_1$ & 500 $\,\mathrm{m}$  \\
 $h_2$ & 2500 $\,\mathrm{m}$  \\
 $N$ & 9.54 $\times 10^{-4} \, \mathrm{s}^{-1}$ \\
 $\lambda$ &  $5 \times 10^{-7} \, s^{-1}$ \\
\hline
\end{tabular}
\end{center}
\end{table}

\begin{figure*}
    \centering
    \includegraphics[width = 39pc]{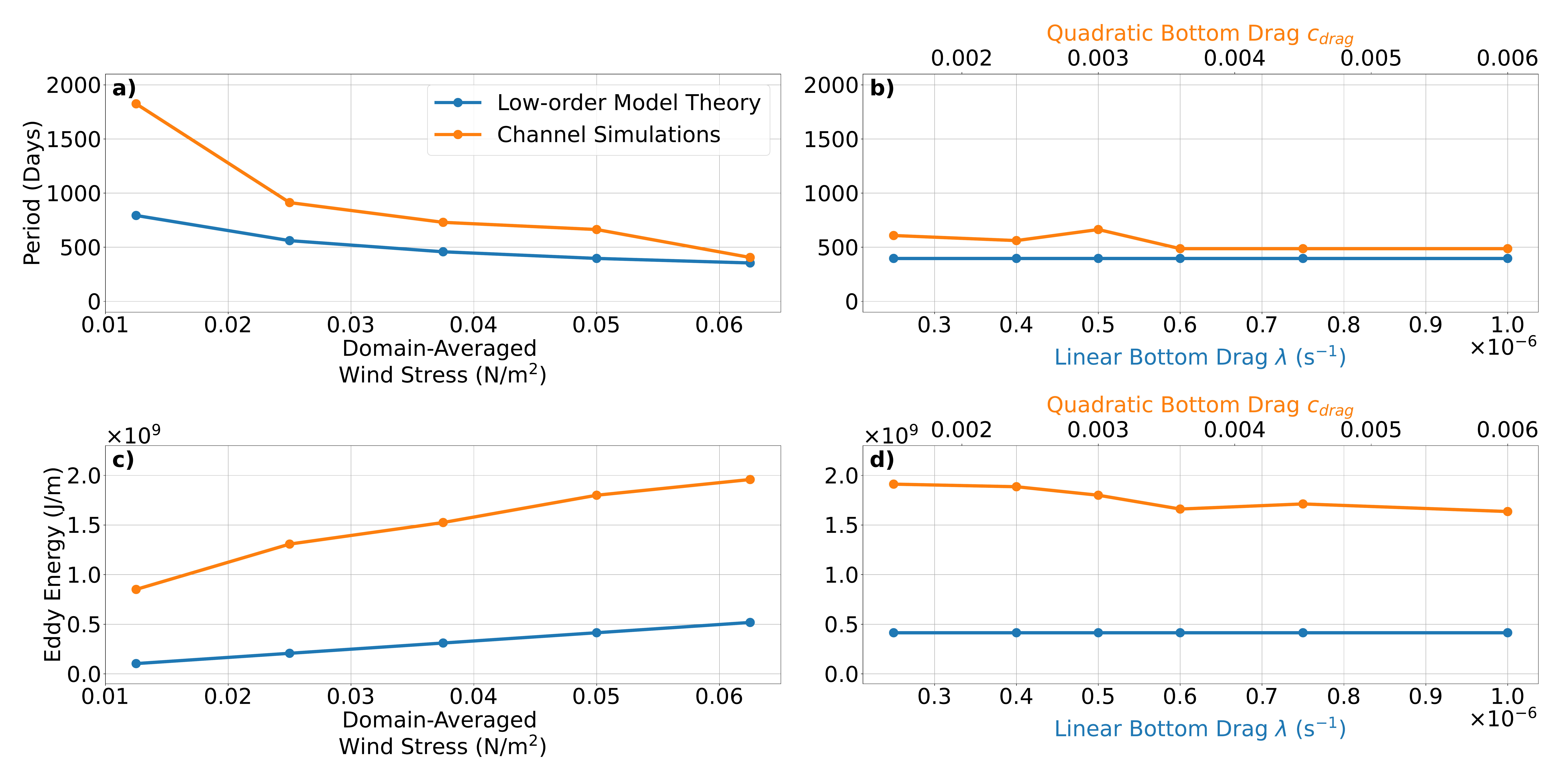}
    \caption{Period of oscillation in $E$ from the channel model simulations and in the low-order theory, with varying wind stress (a) and bottom drag (b).  The dominant frequency for each of the channel model simulations was selected from Fourier spectral analysis and its period is plotted in orange, while the low-order model~(\ref{eqn:period}) used values from the channel model simulations to predict the period of oscillation (see Table~\ref{t1}).  Magnitude of equilibrium Total Eddy Energy from the channel model simulations and in the low-order theory, with varying wind stress (c) and bottom drag (d).}
    \label{fig:oscillatoryslnperiod}
\end{figure*}
\begin{figure*}
    \centering
    \includegraphics[width = 27pc]{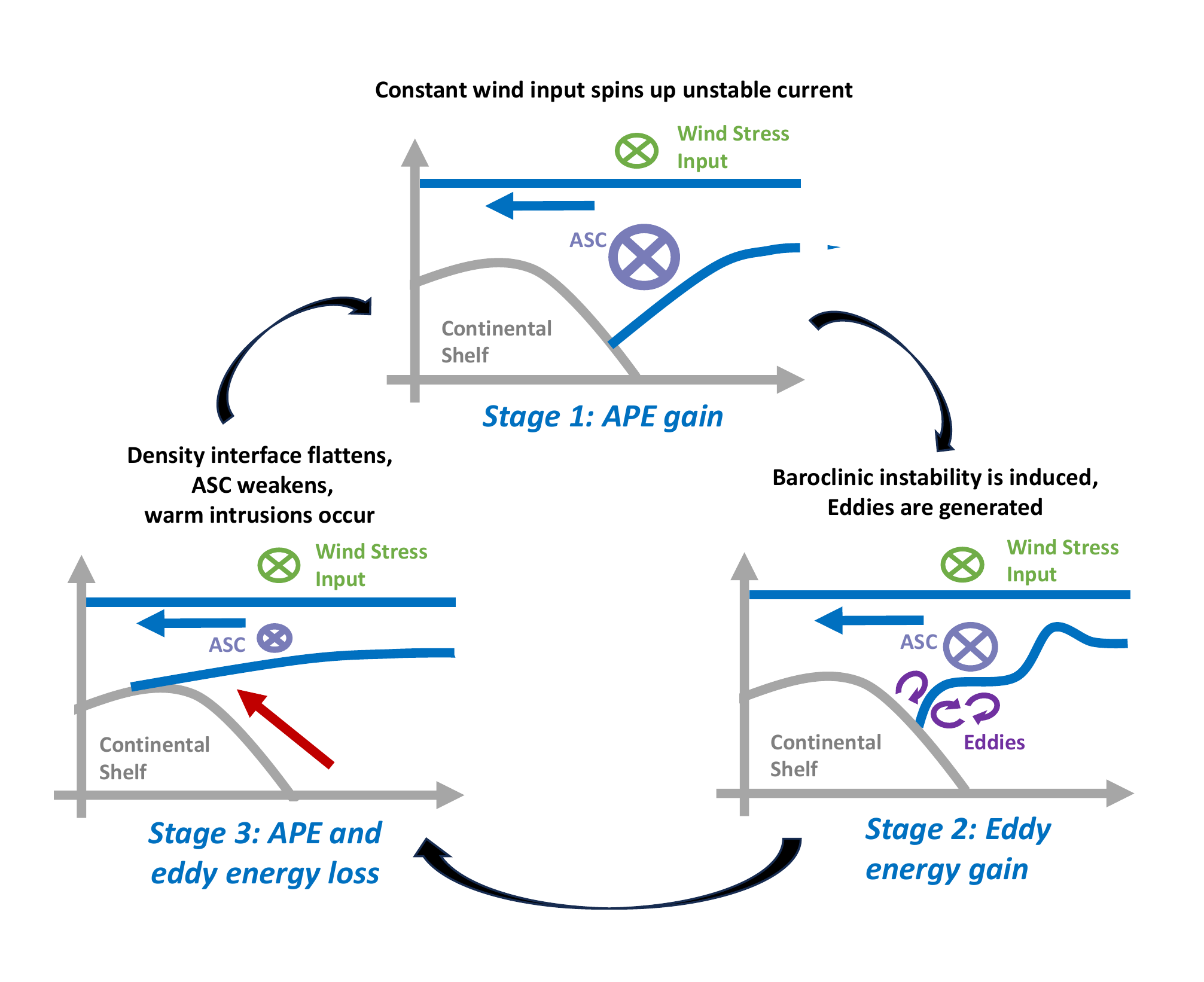}
    \caption{Schematic illustrating the intrinsic oscillation of Antarctic Slope Current strength over a canyon and its governing mechanisms. Stage 1: the constant wind stress input (green) spins up an unstable ASC (purple) through the Ekman transport of surface water and the steepening of isopycnals. Hence, available potential energy (APE) is gained in the current system. Stage 2: baroclinic instability generates eddies and resulting in a gain in eddy energy. Stage 3: the eddies flatten the density interface and weaken, slowing the ASC and shoaling the density surface on the continental shelf. A pathway for warm CDW to access the continental shelf is provided (red arrow), with the Ekman transport of surface water kept constant throughout the cycle (blue arrow) }
    \label{fig:schematicASCoscillation}
\end{figure*}
According to linear stability analysis of the low-order model equilibrium, the oscillation period is expected to decrease as wind stress is increased, but is independent of bottom drag. Fourier spectral analysis was conducted on the eddy kinetic energy time series in each channel model simulation. The dominant frequency was selected and its period is plotted in orange for each of the channel model simulations in Figure~\ref{fig:oscillatoryslnperiod}, with (a) for simulations varying wind stress, and (b) for simulations varying bottom drag. The period predicted by the low-order model~\eqref{eqn:period} is shown in blue, and we have substituted values from the channel model simulations to estimate the period and conduct the parameter sensitivity tests (see Table~\ref{t1}). We find that the period in the channel model simulations decreases as the strength of the wind forcing increases, but is independent of bottom drag, as predicted by the linearisation of the low-order model. The channel model simulations show a period systematically greater than that in the low-order model, with an example shown in Figure~\ref{fig:IVP_mom6} where the control simulation has a longer oscillation period in energy than in the low-order model. Regardless, the trends in the period of oscillation seen in the low-order model still hold in the channel model simulations as wind stress and bottom drag are varied (Figures~\ref{fig:oscillatoryslnperiod}~(a) and~(b)).

In addition to the period of $E$ and $APE$ oscillations, the low-order model parameters can also be used to predict the equilibrium values of $E$ and $APE$ and how they behave as wind stress and bottom drag are varied. From the low-order model, the theoretical equilibrium values of the $E$ and $APE$, $E_0$ and $APE_0$, are proportional to the wind stress input, $\tau_x$, and the square of the bottom drag $\lambda^{2}$ respectively, as in \eqref{eqn:fixedpoint}. The equilibrium energy values in the channel model simulation were found by taking a spatial and time-mean over 20 years of daily data over the continental shelf where the ASC is located, and the comparison between experimental and theoretical $E_0$ as wind and bottom drag are varied is shown in Figures~\ref{fig:oscillatoryslnperiod}~(c) and~(d), with the theoretical equilibrium eddy energy from the low-order model plotted in blue and the channel model simulation results plotted in orange. We see that as $\tau_x$ is increased the $E_0$ for channel model simulations increases in magnitude, matching the prediction by the low-order model. Channel model simulations have a decreasing magnitude of $E_0$ by 10$\%$ over a tripling of bottom drag, while the low-order model predicts that $E_0$ will remain constant.  The equilibrium $APE_0$ in simulations have also been compared with the low-order model prediction (not shown) but the trends in theoretical $APE_0$ and those from the channel simulations do not align.  

These discrepancies in equilibrium energies between the low-order model and channel model simulations are significant and could be due to a number of the assumptions in the low-order model, such as: 1) the channel model simulations use a quadratic drag, while the low-order model uses a linear drag, as addressed earlier in the section; 2) the assumption that the rate of eddy energy generation is directly proportional to eddy energy and vertical shear is an oversimplification, and; 3) the low-order model assumes a flat-bottom channel while the channel model simulations have a continental slope. These could contribute to the difference in trends seen in Figure~\ref{fig:oscillatoryslnperiod}~(d) and equilibrium $APE_0$.  Considering that the effect of the continental slope is excluded from the low-order model, the parallels between the low-order model and the channel model simulations already illustrate key similarities between the systems, especially in the trend in period of oscillation as wind stress and bottom drag are varied.

Comparisons between the low-order model and the channel model simulations indicate that these systems share similar underlying dynamics, allowing us to draw comparisons. Both systems exhibit energy exchanges between $APE$ and $E$, with $APE$ acting as the source of energy for $E$, as enhanced eddy activity acts to flatten the isopycnals. This cycle of $APE$ gain and loss drives the temporal variability of $E$ and the ASC strength. The timescale of this energy exchange is independent of the timescale of external forcings, as the low-order model and the simulated ASC have constant wind forcing applied to the system. The low-order model therefore highlights the intrinsic oscillation in the ASC, where wind forcing and baroclinic instability dominate in turn, as illustrated in Figure~\ref{fig:schematicASCoscillation}. Wind forcing acts to increase the $APE$ of the current system by strengthening the ASC, while baroclinic instability dissipates the $APE$ by converting it into eddy energy, with their equal contribution explaining the oscillatory behavior of the simulated ASC system in the fresh shelf regime.

\section{Discussion and conclusions}
\label{sec:discussion}
We develop an idealised configuration of the ASC in a fresh shelf regime to demonstrate that canyons play a key role in the episodic cross-slope transport of warm CDW water. A cycle of rising and falling rates of eddy generation leads to a temporal variability in ASC strength and eddy kinetic energy in the flow. This temporal variability in eddy energy and ASC strength allows for episodic CDW intrusions on to the continental shelf via canyons, as a weakened current results in an uptilt of isopycnals, facilitating the transport of eddies and CDW onto the continental shelf. We also assess the importance of canyon width on CDW transport in our idealised channel simulations with a constant wind forcing; wider canyons allow for a greater transport of CDW onto the shelf with distinct episodes of intrusion at a regular frequency, while narrower canyons result in greater asymmetry between the draining and isolated pulses of warm CDW intrusions onto the shelf.

We additionally find that the temporal variability of CDW intrusions and the ASC is intrinsic to the idealised ASC system, as it is present even in this model with constant external forcings. The intrinsic variability exists as a consequence of the feedbacks between wind energy input and eddy generation. This relationship is supported by a low-order model of the ASC, which exhibits similar behaviour to the isopycnal channel model simulations. In the low-order model, an intrinsic temporal variability in the current system is set up, as a regular cycle of energy gain and loss occurs. Available potential energy is converted to eddy energy and dissipated as the ASC becomes baroclinically unstable, but potential energy is generated as wind forcing continues to act on the system and spins up a strong current again, restarting the cycle, as illustrated in Figure~\ref{fig:schematicASCoscillation}. The result that baroclinic instability is the dominant cause of variability in the modelled ASC comes with a caveat; this is an idealised model with constant forcings and thus does not impose any variability or other water mass transformations. Hence, baroclinic instability may play a different role in the variability of CDW intrusions seen in more realistic models and in observations. 

There are currently few realistic modelling studies or observations of the ASC in the East Antarctic, which limits our understanding of the intrinsic variability in the ASC system. In regional models and observations, the presence of temporal variability across a range of timescales, irregular topography, and time-varying external forcings makes it difficult to isolate a dominant frequency of intrinsic oscillation in ASC strength. Additionally, both the surface-intensified and bottom-intensified slope current are present in certain dense shelf regions of the East Antarctic, such as the Prydz Bay and downstream region. Therefore, our proposed mechanism driving an intrinsic variability in CDW intrusions and the ASC may not be directly applicable to these East Antarctic regions, as our isopycnal channel model and the low-order model are unable to simulate the bottom-intensified slope currents linked to dense shelf water export. However, further work could apply the physical understanding of the ASC, gained from using the low-order model, to realistic models and observations of the fresh shelf regions and diagnose the intrinsic variability that we see in the channel model simulations. New observation programs could also target longer periods of continuous measurements, such that interannual variability in the ASC and CDW intrusions could be captured in observations. Observing and understanding this intrinsic variability can significantly improve our knowledge of how CDW intrusions  in fresh shelf regions  occur, which in turn would be important for predicting basal melt rates on the Antarctic continental shelf. 

Although our idealised isopycnal channel simulations and low-order model capture the basic structure of the fresh shelf regime and the intrinsic oscillation of the ASC, there exist limitations to our study. For example, investigating the effect of varying canyon topography on cross-shelf dynamics, including the effects of canyon length, steepness of canyon slope, and continental slope \citep[e.g.][]{Isachsen2011BaroclinicSimulations,Zhang2011ShelfTopography,Stern2015InstabilityBreak,Bai2021DoesCurrents, Si2022CoupledRedistribution}, was outside the scope of this study. The isopycnal channel simulation does not include sea ice, tides, nor the effects of water mass transformations \citep{Stewart2019AccelerationTides,Si2022CoupledRedistribution,Si2023HeatGradients}. Hence, this isopycnal configuration cannot be used to address issues such as the effect of a freshened shelf via additional meltwater on the ASC \citep{Moorman2020ThermalModel}. Given the projected weakening of the coastal easterlies \citep{Neme2022ProjectedMargin}, assessing the impact of wind strength changes on CDW intrusions around the East Antarctic using this isopycnal channel simulation is also an important area of future work. Additionally, the low-order model does not include a coupled effect between stratification and current width, where in channel model simulations a narrower current is formed as stratification is reduced (not shown). Hence, to predict the period of the oscillation, the stratification and current width variables in the low-order model prediction in Equation~\eqref{eqn:period} both have to be modified as there are other compensating mechanisms present. The criterion required for this ASC oscillation to occur has also not been determined, but previous work in similar setups have pointed to the existence of a critical threshold for which periodic behaviour in a jet can occur \citep{Hogg2006InterdecadalOcean, Chekroun2022TransitionsModel}. Understanding the criteria required and the experimental parameters key to inducing this oscillatory behaviour in the ASC is a compelling area of future research, and could also be applied to other boundary current systems around the globe.


Our idealised representation of the ASC captures a previously unknown feature of CDW intrusions and ASC strength: namely, their potential intrinsic temporal variability and the direct causal link between ASC weakening and CDW intruding onto the continental shelf, via baroclinic instability and eddy generation. We conclude that a time-varying external forcing is not required to force interannual variability in ASC strength and CDW intrusions. There could thus be additional variability in CDW intrusions not captured in non-eddy resolving models. Further work could assess the conditions required for the intrinsic variability to occur and diagnose this variability in realistic models and observations in an effort to improve our understanding of CDW intrusions around East Antarctica.

\acknowledgments
This research was supported by the Australian Research Council Special Research Initiative, Australian Centre for Excellence in Antarctic Science (ARC Project Number SR200100008). E.Q.Y.O.~is supported by the Australian Government Research Training Program Scholarship (RTP). N.C.C.~is supported by the Australian Research Council DECRA Fellowship DE210100749. A.McC.H. and M.H.E.~(DP190100494) acknowledge funding from the Australian Research Council. This project received grant funding from the Australian Government as part of the Antarctic Science Collaboration Initiative program (ASCI000002). Computational resources were provided by the Australian National Computational Infrastructure at the ANU, which is supported by the Commonwealth Government of Australia. We thank the three anonymous reviewers for their constructive feedback that greatly improved the manuscript.

%
%
\datastatement
Scripts used for analysis and for reproducing figures will be available at \texttt{github.com/ongqingyee/idealised-ASC} upon acceptance of this manuscript. The source code for the MOM6 simulation run is available on \texttt{github.com/mom-ocean/MOM6}. Simulation output used to reproduce figures will be available in a Zenodo repository upon acceptance of the manuscript.

%






%
%
%

\appendix
\appendixtitle{Derivation of Coupled Ordinary Differential Equations for Total Eddy Energy and Available Potential Energy in Low-Order Model}
\label{appen:eddyeqn_derivation}
We express the eddy budget in~\eqref{eqn:word_Eeqn} and~\eqref{eqn:word_APEeqn} in terms of the depth-integrated total available potential energy per unit longitudinal length, $APE$, and the depth-integrated total eddy energy per unit longitudinal length, $E$. We first define $APE$ in this two-layered model; see Figure~\ref{fig:2layer_energy_ODE_setup}. We have one free interface between the two fluid layers, $\eta(y)$, a rigid lid and define the thickness of each layer as $h_i$. We assume the slope of the density interface is constant within the model domain, such that $\eta = \sgn(\tau_x) \sgn(f) s y$, where $s > 0$ is the magnitude of the slope of the density interface. The sign of the slope, $\partial \eta/\partial y$, depends on the sign of the wind stress and the Coriolis parameter. In a domain of latitudinal length $L_y$, over which the slope of the density surface is constant, the $APE$ is:

\begin{align}
    APE & = \int^{L_y}_0 \frac{1}{2} \rho_0 g' \eta^2 \, \mathrm{d}y \nonumber\\
    & = \frac{1}{6}\rho_0 g' s^{2}L_y^{3},
    \label{eq:app_APE_slope}
\end{align}

\noindent where $\rho_0$ is the reference density and $g'=g\Delta \rho / \rho_0$ is reduced gravity of the two-layer system. 

Defining the eddy energy conversion term as $ \vert f \vert/N \, |\partial u/\partial z| \, E$ \citep{Marshall2012AFluxes, Marshall2017EddyCurrent}, where $E = \sum_i \int \large( \tfrac1{2} \rho_0 h_i |\mathbf{u}''_i|^2 +  \tfrac1{2} \rho_0 g' \eta''^2 \large) \mathrm{d} y$, we can find an expression for this term using the parameters of the layered model. We can also use that for a two-layer model $N^2 = 2g'/H$. We approximate the vertical shear $\partial u/\partial z$ discretely as $\partial u /\partial z \approx (u_1 - u_2) / (H/2)$. Furthermore, by assuming constant linear shear and also zero bottom velocity $u_{\rm bot} = 0$, we can integrate vertical shear up to get that the surface velocity is $u_{\rm surface} = H \partial u/\partial z \approx 2(u_1 - u_2)$. To remove the velocity terms in each layer, we use the thermal wind balance $f (u_1- u_2) = g' \partial \eta/\partial y = \sgn(\tau_x) \sgn(f) g's$ and combine all the terms above to write the expressions for the rate of change of depth-integrated eddy energy in terms of $APE$ to arrive at~\eqref{eqn:E_eqn}.

Next, we define the rate of change of $APE$, the energy input from winds as $\iint \boldsymbol{\tau} \cdot \mathbf{u}_{\rm surface} \, \mathrm{d}A$. If we consider only the zonal flow component, the work per unit longitude is:
\begin{equation}
    \textrm{Wind Input} = \int \tau_x u_{\rm surface} \, \mathrm{d}y .
    \label{eq:app_wind_input}
\end{equation}

To express the wind energy input in terms of $APE$ and $E$, we use the surface velocity $u_{\rm surface} = 2(u_1 - u_2)$ and find the zonal velocities in individual layers through geostrophy to express the wind energy input in terms of the slope of the density interface:
\begin{align}
    u_{1g}  = & \frac{-1}{\rho_0 f}\frac{\partial p_{\rm atm}}{\partial y}, \\
    u_{2g}  = & -\frac{1}{\rho_0 f} \left( \frac{\partial p_{\rm atm}}{\partial y} + g' \rho_0 \frac{\partial \eta}{\partial y} \right) ,
\end{align}
which when combined give:

\begin{equation}
    u_{\rm surface} = 2(u_{1g} - u_{2g}) = \sgn(\tau_x) \frac{2g's}{|f|} .
\end{equation}

Substituting the expression for $u_{\rm surface}$ into~\eqref{eq:app_wind_input}, using~\eqref{eq:app_APE_slope} to write $s$ in terms of $APE$, and using the same eddy energy conversion term as for \eqref{eqn:E_eqn}, we end up with~\eqref{eqn:APE_eqn}.


\end{document}